\documentclass[twocolumn,prb,amsmath,amssymb,superscriptaddress,showpacs]{revtex4}
\usepackage{graphicx}
\usepackage{bm}
\usepackage{bbm}
\usepackage{epstopdf}
\usepackage{mathbbol}
\usepackage{color}
\newcommand{\Tr}{\text{Tr}}
\newcommand{\EZ}{E_{Z}}
\newcommand{\ket}[1]{\left|#1\right>}
\newcommand{\bra}[1]{\left< #1 \right|}

\newcommand{\MA}{\mathcal{A}}
\newcommand{\ua}{\uparrow}
\newcommand{\da}{\downarrow}
\begin{document}
\title{Decoherence of an exchange qubit by hyperfine interaction}
\author{Jo-Tzu Hung}
	\affiliation{Department of Physics, University at Buffalo, SUNY, Buffalo, New York 14260-1500, USA}
\author{Jianjia Fei}
	\affiliation{Department of Physics, University of Wisconsin-Madison, Madison, Wisconsin 53706, USA}
\author{Mark Friesen}
	\affiliation{Department of Physics, University of Wisconsin-Madison, Madison, Wisconsin 53706, USA}
\author{Xuedong Hu}
	\affiliation{Department of Physics, University at Buffalo, SUNY, Buffalo, New York 14260-1500, USA}
\date{\today}
\begin{abstract}
	We study three-electron-spin decoherence in a semiconductor triple quantum dot with a linear geometry. The three electron spins are coupled by exchange interactions $J_{12}$ and $J_{23}$, and we clarify inhomogeneous and homogeneous dephasing dynamics for a logical qubit encoded in the ($S\!=\!1/2,S^{z}\!=\!1/2$) subspace. We first justify that qubit leakage via the fluctuating Overhauser field can be effectively suppressed by sufficiently large Zeeman and exchange splittings. In both $J_{12}\!=\!J_{23}$ and $J_{12}\!\neq\! J_{23}$ regimes, we construct an effective pure dephasing Hamiltonian with the Zeeman splitting $\EZ \gg J_{12},J_{23}$. Both effective Hamiltonians have the same order of magnitude as that for a single-spin qubit, and the relevant dephasing time scales ($T_2^*$, $T^{FID}$, etc.) are of the same order as those for a single spin.  We provide estimates of the dynamics of three-spin free induction decay, the decay of a Hahn spin echo, and the decay of echoes from a CPMG pulse sequence for GaAs quantum dots.
\end{abstract}
\pacs{03.67.Lx; 73.21.La; 03.65.Yz; 85.35.Be}
\maketitle

\section{Introduction}\label{sec:introduction}

Confined electron spins are promising candidates as qubits in a solid state quantum computer.  In the past decade there has been tremendous progress in the theoretical and experimental studies of single and two electron spin states.\cite{Hanson_RMP07}  Coherence measurement and control has been performed in single-electron quantum dots (QDs).\cite{Koppens_PRL08,Tyryshkin_NatMat11,Morello_Nature10} Two-electron-spin manipulation and coherence control have also been demonstrated.\cite{Petta_Science05, Bluhm_NP11, Maune_Nature12}
Theoretically, it has been established that hyperfine (hf) interaction induced pure dephasing is the main source of decoherence for a single-spin qubit in GaAs or natural Si, whether confined by a QD or a donor.\cite{Cywinski_PRB09} For two exchange-coupled spins in a double QD, it has also been shown that hf interactions and charge fluctuations contribute significantly to the decoherence,\cite{Coish_PRB05, Hu_PRL06, Witzel_PRB08, Yang_PRB08a, Hung_PRB13, Dial_PRL13} while other decoherence channels could potentially be relevant as well.\cite{Prada_PRB08, Roszak_PRB09, Borhani_PRB10, Hu_PRB11, Raith_PRL12}

\begin{figure}[t]
	\includegraphics[width=0.5\linewidth]{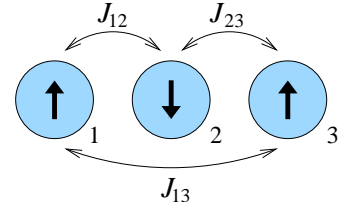}
	\caption{Schematics of a linear triple dot, with one electron per dot.  The interdot spin couplings are of the Heisenberg exchange type, labeled by $J_{12}$, $J_{23}$, and $J_{13}$. In several recent experimental implementations,\cite{Laird_PRB10,Medford_NatNano13, Medford_PRL13} $J_{13}\simeq0$. }\label{Fig:3dot}
\end{figure}

Recently, studies of three-electron-spin dynamics in a double or triple QD have been attracting wide attention both experimentally\cite{Gaudreau_PRL06, Gaudreau_PRB09, Laird_PRB10, Gaudreau_NP12, Shi_PRL12, Medford_NatNano13, Medford_PRL13} and theoretically.\cite{Byrd_PRA05, Korkusinski_PRB07, Hsieh_PRB12b, Koh_PRL12, Ladd_PRB12, Troiani_PRB12, Taylor_PRL13, Mehl_PRB13, Doherty_PRL13}  One particularly interesting problem is the three-spin encoding for a logical qubit in a linear triple QD,\cite{DiVincenzo_Nature00, Medford_PRL13} as illustrated in Fig.~\ref{Fig:3dot}.  The two states of the logical qubit both have $S = 1/2$ and $S^{z} =1/2$, so that they form a decoherence-free subspace against noise in a uniform external magnetic field.  Best of all, both single-qubit and two-qubit operations of these logical qubits can be fully controlled electrically, via exchange interactions between neighboring dots.\cite{Taylor_PRL13,Doherty_PRL13}

There are two major sources of decoherence for three exchange-coupled spins: magnetic noise and charge noise.  Generally, a three-electron-spin state always has a finite spin density in space, similar to single-spin qubits (but in contrast to $S\!-\!T_0$ qubits at the finite-$J$ limit\cite{Hung_PRB13}).  Therefore, nuclear spin induced pure dephasing, which is the dominant decoherence mechanism for single spins,\cite{Cywinski_PRB09,Cywinski_PRL09} should also be an important (if not the dominant) source of decoherence for three-electron-spin states.  When exchange interactions are turned on, charge noise due to electrical fluctuations and/or phonons can also dephase the logical qubit.\cite{Coish_PRB05, Hu_PRL06, Hu_PRB11}  However, in the present paper we will not explore the effect of charge noise beyond a qualitative discussion, and will instead focus on decoherence of the logical qubit in a triple QD due to hf interactions with the surrounding nuclear spins.

In this paper, we consider the all-exchange qubit\cite{DiVincenzo_Nature00} operated in the $J$-always-on limit.\cite{Medford_PRL13}  This qubit is coded with $\ket{g}\!=\! \frac{1}{\sqrt{3}} \ket{T_{0}}_{13} \ket{\uparrow}_{2} - \sqrt{\frac{2}{3}} \ket{\uparrow\uparrow}_{13} \ket{\downarrow}_{2}$ and $\ket{e}\!=\!\ket{S}_{13} \ket{\uparrow}_{2}$, where $\ket{S}_{13}$($\ket{T_{0}}_{13}$) represents an unpolarized singlet (triplet) formed by spins $1$ and $3$. We clarify the relevant system-reservoir Hamiltonian, and identify possible leakage, relaxation, and dephasing channels.  After discussing how to suppress qubit leakage, we calculate both inhomogeneous broadening and homogeneous broadening of the logical qubit, as well as spin echo decay, and compare our results with the decoherence of single-spin qubits. We note that this choice of the three-spin qubit is slightly different from the original proposal in Ref.~\onlinecite{DiVincenzo_Nature00}, where the roles of spins 2 and 3 are switched.  While this switch is an important change when considering initialization and manipulation of the qubit,\cite{Medford_PRL13} it is not significant from the perspective of projecting the hf interaction onto the state basis.  Therefore the calculations done in this work, and the qualitative behavior we discuss, are applicable to both variants of the encoding scheme as long as the exchange couplings are always on.~\cite{DiVincenzo_Nature00, Medford_PRL13}

The original proposal\cite{DiVincenzo_Nature00} considers qubits encoded as $\ket{S}_{12}\ket{\ua}_{3}$ and $\sqrt{\frac{2}{3}} \ket{\ua\ua}_{12} \ket{\da}_{3} - \frac{1}{\sqrt{3}} \ket{T_{0}}_{12} \ket{\ua}_{3}$, which correspond to eigenstates in the limit $J_{23} \ll J_{12}$.
For this encoding scheme, the leading-order effects of the hyperfine interaction, associated with the longitudinal Overhauser field, have previously been analyzed in two different limits:\cite{Ladd_PRB12} (1)  when the exchange between the qubit states is large compared to the Overhauser fluctuations, the oscillating $\ket{S}_{12}\ket{\ua}_{3}$ signal in Rabi experiments includes a $T^{*}_{2}$-type decay that depends only on the variations of the Overhauser field in the three dots; (2) in the opposite limit where the exchange splitting vanishes, the oscillations decay similarly to those in a double QD with a vanishing exchange splitting.\cite{Coish_PRB05,Petta_Science05} In contrast to Ref.~\onlinecite{Ladd_PRB12}, we focus here on the regime where both $J_{12}$ and $J_{23}$ are always on.  As shown below, leakage from the logical qubit Hilbert space due to hf interaction can then be suppressed by requiring $J_{12}$ and $J_{23}$ to be much larger than the Overhauser fluctuations.

Our work is organized as follows. In Sec.~\ref{sec:3spins}, we present the model Hamiltonian for three electron spins in a triple QD. We characterize the spin eigenstates and their properties in Sec.~\ref{sec:states}, and express the hf interactions in the relevant eigenbasis in Sec.~\ref{sec:Hhf}. In Sec.~\ref{sec:decoherence}, we study the decoherence due to hf interactions by deriving an effective pure dephasing Hamiltonian in the $J_{12}=J_{23}$ regime, and we calculate the free induction decay (FID) and other coherence decays after applying Hahn echo (HE) or 2-pulse CPMG sequences to GaAs QDs. In Sec.~\ref{sec:nonuniform}, we derive an effective Hamiltonian in the $J_{12}\neq J_{23}$ regime, and we calculate the FID for GaAs QDs. Our conclusions are presented in Sec.~\ref{sec:conclusion}.

\section{Three electron spins in a triple quantum dot}\label{sec:3spins}
The logical qubits we consider are the eigenstates of the linear, three-spin chain shown in Fig.~\ref{Fig:3dot}, with uniform exchange couplings between just the nearest neighbors, such that $J_{12}\!=\!J_{23}\!=\!J$ and $J_{13}\!=\!0$.  This simple scenario corresponds exactly to the resonant exchange qubit.\cite{Medford_PRL13}  A uniform magnetic field is generally also applied to the triple dot.

The model Hamiltonian includes the terms $\hat{H}= \hat{H}_{\text{e}} + \hat{H}_{\text{N}} + \hat{H}_{\text{hf}}$.  Here $\hat{H}_{\text{e}}$ is the electronic Hamiltonian, including Zeeman and exchange interactions, and $\hat{H}_{\text{N}}$ is the nuclear Zeeman energy:
\begin{eqnarray}
	\hat{H}_{\text{e}} &=& \EZ \sum^{3}_{d=1} S^{z}_{d}+
		J\left({\bf S}_1 \cdot {\bf S}_2 + {\bf S}_2 \cdot {\bf S}_3 \right) + \Delta{\bf S}_2 \cdot {\bf S}_3,\quad\label{eq:He}\\
	\hat{H}_{\text{N}} &=& \sum_{i}\omega_{i[\alpha]} I^{z}_{i}\label{eq:HN}\;.
\end{eqnarray}
where we have identified electrons 1 through 3 as the spins residing in dots $d= 1, 2, 3$, respectively, and $\mathbf{S}_{d}$ ($\mathbf{I}_{i}$) are the spin operators for the $d^{th}$ ($i^{th}$) electron (nuclear) spin. Here, $\EZ=g\mu_{B}B$ is the Zeeman splitting for the electron spins, and $\omega_{i[\alpha]}$ is the $i^{th}$ nuclear spin of species $\alpha$.
We have also introduced the parameters $J\equiv J_{12}$ and $\Delta\equiv J_{23}-J_{12}$, that allow us to smoothly vary between a uniform chain, where $J_{12}=J_{23}$ and $\Delta=0$, and a nonuniform chain, where $J_{12}\neq J_{23}$ and $\Delta\neq 0$.  For the contact hf interaction $\hat{H}_{\text{hf}}$, we do not include the effects of inter-dot orbital overlaps,\cite{Hung_PRB13} since they are relatively weak when the overlaps are small, as in the case we consider.  Within this approximation, we have
\begin{eqnarray}
	\hat{H}_{\text{hf}}&\approx&\sum_{d=1}^{3}\sum_{i\in d} A_{i}\left( S^{z}_{d}I^{z}_{i}+
		\frac{ S^{+}_{d}I^{-}_{i}+S^{-}_{d}I^{+}_{i}}{2}\right)\nonumber\\
		&=& \sum_{d=1}^{3} \mathbf{S}_{d}\cdot \mathbf{B}_{d}\;,\quad
		\mathbf{B}_{d}= \sum_{i \in d} A_{i}\mathbf{I}_{i}\;. \label{eq:HF_approx}
\end{eqnarray}
Here $A_{i\in d}$ is the hf coupling between the electron and the $i^{th}$ nuclear spin in dot $d$; the electron spin in dot $d$ then experiences the local nuclear Overhauser field $\mathbf{B}_{d}$.  Under typical experimental conditions,\cite{Hanson_RMP07} the nuclear spins are randomly oriented, so that the average nuclear polarization vanishes $\langle \mathbf{B}_{d} \rangle = 0$, while $\sqrt{\langle\mathbf{B}^{2}_{d}\rangle} \sim \MA/\sqrt{N}$, where $\MA$ and $N$ are the total hf energy and the number of unit cells in dot $d$.

In the remainder of this section, we characterize the energy spectrum of the spin system, and we express $\hat{H}_{\text{hf}}$ in the basis of Table~\ref{table:coupled_spin_states}.

\begin{table}[t]
\begin{tabular}{lccccc}
\hline
\hline & $\Psi$ & $E(J,E_Z)$ & $S$ & $S^{z}$ & $S_{13}$  \\
\hline
$D^{\prime}_{1/2}$ & $\frac{1}{\sqrt{6}} \left( \ket{\da \ua \ua} -2 \ket{\ua \da \ua} + \ket{\ua\ua \da} \right)$
	& $-J +\frac{1}{2} E_Z$ & $\frac{1}{2}$ & $\frac{1}{2}$ & 1 \\
$D^{\prime}_{-1/2}$ & $\frac{1}{\sqrt{6}} \left( \ket{\ua \da \da}- 2 \ket{\da \ua \da}+ \ket{\da \da \ua} \right)$
	& $-J - \frac{1}{2} E_Z$ & $\frac{1}{2}$ & $-\frac{1}{2}$ & 1  \\
$D_{1/2}$ & $\frac{1}{\sqrt{2}} \left( \ket{\da \ua \ua}- \ket{\ua\ua \da} \right)$ & $+\frac{1}{2} E_Z$ &
	$\frac{1}{2}$ & $\frac{1}{2}$ & 0 \\
$D_{-1/2}$ & $\frac{1}{\sqrt{2}} \left( \ket{\ua\da\da}-\ket{\da\da\ua} \right)$ & $-\frac{1}{2} E_Z$ &
	$\frac{1}{2}$ & $-\frac{1}{2}$ & 0 \\
$Q_{3/2}$ & $\ket{\ua\ua\ua}$ & $\frac{1}{2}J +\frac{3}{2} E_Z$ & $\frac{3}{2}$ & $\frac{3}{2}$ & 1  \\
$Q_{1/2}$ & $\frac{1}{\sqrt{3}} \left( \ket{\da \ua\ua} + \ket{\ua\da \ua} + \ket{\ua \ua \da} \right)$ & $\frac{1}{2}J+\frac{1}{2} E_Z$
	& $\frac{3}{2}$ & $\frac{1}{2}$ & 1 \\
$Q_{-1/2}$ & $\frac{1}{\sqrt{3}} \left( \ket{\ua\da \da}+ \ket{\da\ua\da}+\ket{\da\da\ua} \right)$ & $\frac{1}{2}J - \frac{1}{2} E_Z$ & $\frac{3}{2}$ & $-\frac{1}{2}$ & 1 \\
$Q_{-3/2}$ & $\ket{\da\da\da}$ & $\frac{1}{2}J - \frac{3}{2} E_Z$ &
	$\frac{3}{2}$ & $-\frac{3}{2}$ & 1 \\
\hline
\hline
\end{tabular}
\caption{Three-spin eigenstates with spin quantum numbers and their energies for a uniformly coupled triple-dot chain. The first two columns are the state labels and the states themselves. The third column gives the energy of the state. Columns four to six describe the spin quantum numbers, where $S$ and $S^z$ refer to the total spin and total spin in the $z$ direction for all three electrons, while $S_{13}$ refers to the total spin for the two electrons in dots 1 and 3. For example, $D^{\prime}_{-1/2}$ represents the $S^{z}=-1/2$ state in the doublet $D^{\prime}$.} \label{table:coupled_spin_states}
\end{table}
\subsection{Three-electron-spin states in a linear triple QD}\label{sec:states}
\begin{figure}[h]
	\includegraphics[width=0.8\linewidth]{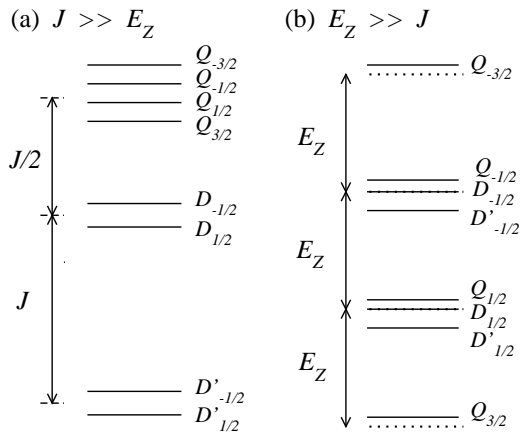}
	\caption{Energy spectrum of three linearly coupled electron spins in a uniform magnetic field, following the labeling scheme indicated in Table I, for two limiting cases:  (a) $J \gg E_Z$, and (b) $E_Z \gg J$. For GaAs quantum dots, with a negative $g$-factor ($g$=$-0.44$), $\ket{Q_{3/2}}$ is the ground state.}\label{fig:3spinSpectrumCoupled}
\end{figure}

We start by discussing the three-spin electronic energy spectrum of a uniform chain with $\Delta=0$.  In this case, the electronic Hamiltonian $\hat{H}_{e}$ simplifies to
\begin{equation}
	\hat{H}_{\text{e}}(\Delta=0) = \EZ S^{z} + \frac{J}{2} \left[ S^{2} - S_{13}^2 - \frac{3}{4} \right] \,,\label{eq:He0}
\end{equation}
where $S$ and $S^{z}=\sum^{3}_{d=1}S^{z}_{d}$ are the total spin quantum numbers for all three electrons, whereas  $S_{13}$ is the total spin quantum number for electrons $1$ and $3$.  From Eq.~(\ref{eq:He0}), we see that the eigenstates of $\hat{H}_{\text{e}}(\Delta=0)$ are simultaneous eigenstates of $S^{2}$, $S_{13}^{2}$, and $S^{z}$, comprising a quadruplet ($Q$) with $S\!=\!3/2$ and two doublets ($D$ and $D^{'}$) with $S\!=\!1/2$. In Table~\ref{table:coupled_spin_states}, we detail all the eigenstates with their state configurations. Figure~\ref{fig:3spinSpectrumCoupled} shows the corresponding energy spectrum at low magnetic fields ($E_Z \ll J$), which splits into three manifolds: $D^{'}$, $D$ and $Q$. In this regime, the triple dot can act as a spin bus,\cite{Friesen_PRL07} with the ground doublet acting as a effective spin-1/2 system.  For high magnetic fields ($E_Z \gg J$), the spectrum splits into four manifolds defined by $S^z$, separated roughly by the Zeeman splitting.  Considering the typical experimental parameters $B = 1$~T (corresponding to $|E_Z| \simeq 25$~$\mu$eV in GaAs), and $J = 1$-$10$~$\mu$eV, we see that the $E_{Z}\gg J$ regime suggests encoding our logical qubits in the states $|g\rangle = \ket{D^{\prime}_{1/2}}$ and $|e\rangle = \ket{D_{1/2}}$, which have the same $S$ and $S^z$ quantum numbers.

We note that $|g\rangle$ and $|e\rangle$ only represent eigenstates of the uniform ($\Delta=0$) three-spin chain. When $\Delta$ is nonzero in Eq.~(\ref{eq:HN}), states $|g\rangle$ and $|e\rangle$ are no longer eigenstates; however, $\left[ {\bf S}_2 \cdot {\bf S}_3, S^z \right] = 0$, so $S^z$ is still a good quantum number, and the four Zeeman manifolds in Fig.~\ref{fig:3spinSpectrumCoupled}(b) remain uncoupled.  It is therefore practical to adopt $|g\rangle$ and $|e\rangle$ as a basis for the logical qubit, regardless of $\Delta$. We therefore focus on the $S^z = 1/2$ manifold, comprised of states $\ket{D^{\prime}_{1/2}}$, $\ket{D_{1/2}}$, and $\ket{Q_{1/2}}$.  The electronic Hamiltonian in this subspace is given by
\begin{equation}
	\hat{H}_{\text{e},\Delta}^{S^{z}=1/2} = \frac{\EZ}{2}\mathbb{1} +\left(\begin{array}{ccc}
			-J-\frac{1}{2} \Delta & \frac{\sqrt{3}}{4} \Delta & 0 \\
			\frac{\sqrt{3}}{4} \Delta & 0 & 0 \\
			0 & 0 & \frac{J}{2}+\frac{1}{4} \Delta
		\end{array}\right) \label{eq:He_NU}\,,
\end{equation}
where $\mathbb{1}$ is the identity operator. We emphasize that in the analysis so far, the states $\ket{g} = \ket{D^{\prime}_{1/2}}$ and $\ket{e} = \ket{D_{1/2}}$ are not coupled to $Q_{1/2}$.  The only effect of the nonuniform exchange coupling ($\Delta\neq 0$) is to couple $\ket{D^{\prime}_{1/2}}$ and $\ket{D_{1/2}}$. Indeed, Ref.~\onlinecite{Medford_PRL13} uses this fact to achieve $\sigma^x$ rotations for the resonant exchange qubit. We also note that our discussion of the $S^{z}=1/2$ manifold also applies to the $S^{z}=-1/2$ manifold, comprised of states $\ket{D^{\prime}_{-1/2}}$,$\ket{-D_{1/2}}$ and $\ket{Q_{-1/2}}$; the exchange couplings do not distinguish the two manifolds, and their Hamiltonians only differ in the sign of the $E_{Z}$ term. Below we first study the coherence properties of states $|g\rangle$ and $|e\rangle$ when $\Delta = 0$. In Sec.~\ref{sec:nonuniform}, we consider the case of $\Delta \neq 0$.

\subsection{Hyperfine interaction Hamiltonian}\label{sec:Hhf}
The effect of the local and varying Overhauser fields $\mathbf{B}_{d}$ depends on the specific spin state in dots $d=1,2,3$, according to Eq.~(\ref{eq:HF_approx}).
To simplify the evaluation of $\hat{H}_{\text{hf}}$, we introduce the notation
\begin{equation}
	B^{p}_{lmr} = l B_{1}^{p} +  m B_{2}^{p} + r B_{3}^{p} \,, \label{eq:combinations}
\end{equation}
where $p = z, +, -$, and $l,m,r$ can take the values $0,1,2$, or $\bar{1}=-1$.
For example, $\bra{Q_{3/2}}\hat{H}_{hf}\ket{Q_{3/2}}=\bra{\ua\ua\ua}\hat{H}_{\text{hf}}\ket{\ua\ua\ua}=\frac{1}{2}B^z_{111}$, where $B^z_{111}=+B^z_1+B^z_2+B^z_3$.  Equation~(\ref{eq:combinations}) can be viewed as a generalized field gradient, and we will refer to it this way in the following work.
Now $\hat{H}_{\text{hf}}$ can be expressed in the basis of Table~\ref{table:coupled_spin_states}, yielding
\begin{widetext}
\begin{equation}
\hat{H}_{\text{hf}} = \left(\begin{array}{cccccccc}
\frac{1}{6} B_{2\bar{1}2}^z	& -\frac{1}{6} B_{2\bar{1}2}^-	& -\frac{1}{2\sqrt{3}} B_{10\bar{1}}^z
	& \frac{1}{2\sqrt{3}} B_{10\bar{1}}^-	& \frac{1}{2\sqrt{6}} B_{1\bar{2}1}^+	& -\frac{1}{3\sqrt{2}} B_{1\bar{2}1}^z
	& -\frac{1}{6\sqrt{2}} B_{1\bar{2}1}^-	& 0  \\
-\frac{1}{6}B_{2\bar{1}2}^+	& -\frac{1}{6} B_{2\bar{1}2}^z	& \frac{1}{2\sqrt{3}} B_{10\bar{1}}^+
	& \frac{1}{2\sqrt{3}} B_{10\bar{1}}^z	& 0			& -\frac{1}{6\sqrt{2}} B_{1\bar{2}1}^+
	& \frac{1}{3\sqrt{2}} B_{1\bar{2}1}^z & \frac{1}{2\sqrt{6}} B_{1\bar{2}1}^-  \\
-\frac{1}{2\sqrt{3}}B_{10\bar{1}}^z	& \frac{1}{2\sqrt{3}} B_{10\bar{1}}^-	& \frac{1}{2} B_{010}^z
	& -\frac{1}{2} B_{010}^-		& \frac{1}{2\sqrt{2}} B_{10\bar{1}}^+	& -\frac{1}{\sqrt{6}} B_{10\bar{1}}^z
	& -\frac{1}{2\sqrt{6}} B_{10\bar{1}}^- & 0  \\
\frac{1}{2\sqrt{3}} B_{10\bar{1}}^+	& \frac{1}{2\sqrt{3}} B_{10\bar{1}}^z	& -\frac{1}{2} B_{010}^+
	& -\frac{1}{2} B_{010}^z		& 0							& -\frac{1}{2\sqrt{6}} B_{10\bar{1}}^+
	& \frac{1}{\sqrt{6}} B_{10\bar{1}}^z & \frac{1}{2\sqrt{2}} B_{10\bar{1}}^-  \\
\frac{1}{2\sqrt{6}} B_{1\bar{2}1}^-	& 0 							& \frac{1}{2\sqrt{2}} B_{10\bar{1}}^-
	& 0						& \frac{1}{2} {B}_{111}^z			& \frac{1}{2\sqrt{3}} {B}_{111}^-
	& 0 & 0  \\
-\frac{1}{3\sqrt{2}} B_{1\bar{2}1}^z	& -\frac{1}{6\sqrt{2}} B_{1\bar{2}1}^-	& -\frac{1}{\sqrt{6}} B_{10\bar{1}}^z
	& -\frac{1}{2\sqrt{6}} B_{10\bar{1}}^- & \frac{1}{2\sqrt{3}} {B}_{111}^-	& \frac{1}{6} {B}_{111}^z
	& \frac{1}{3} {B}_{111}^- & 0  \\
-\frac{1}{6\sqrt{2}} B_{1\bar{2}1}^+	& \frac{1}{3\sqrt{2}} B_{1\bar{2}1}^z	& -\frac{1}{2\sqrt{6}} B_{10\bar{1}}^+
	& \frac{1}{\sqrt{6}} B_{10\bar{1}}^z	& 0						& \frac{1}{3} {B}_{111}^+
	& -\frac{1}{6} {B}_{111}^z & \frac{1}{2\sqrt{3}} {B}_{111}^-  \\
0	& \frac{1}{2\sqrt{6}} B_{1\bar{2}1}^+	& 0						& \frac{1}{2\sqrt{2}} B_{10\bar{1}}^+
	& 0	& 0	& \frac{1}{2\sqrt{3}} {B}_{111}^+						& -\frac{1}{2} {B}_{111}^z  \\
\end{array}\right) \label{eq:HF_coupled}\;.
\end{equation}
\end{widetext}
We note that the diagonal terms in Eq.~(\ref{eq:HF_coupled}) are nonvanishing because $S^z\neq 0$ for the basis states in Table~\ref{table:coupled_spin_states}.
For the off-diagonal terms, the $S^{z}_{d}$, $S^{+}_{d}$, and $S^{-}_{d}$ terms in Eq.~(\ref{eq:HF_approx}) ensure couplings between states with $\Delta S^z=0,1$; all other couplings vanish.
Of particular interest, states within a given $S^{z}$ manifold are coupled by Overhauser field gradients, which are similar to, but more complex than, what has been studied for $S$-$T_{0}$ qubits in double QDs.\cite{Coish_PRB05,Yang_PRB08a,Hung_PRB13}

\section{Decoherence due to hyperfine interactions}\label{sec:decoherence}

We now study decoherence between the logical qubit states $\ket{g} = \ket{D^{\prime}_{1/2}}$ and $\ket{e} = \ket{D_{1/2}}$.  First we establish the optimal operating regime of the logical qubit in which qubit leakage is minimized, then we construct an effective Hamiltonian within the logical qubit Hilbert space, and calculate decoherence within this subspace.

\subsection{Qubit leakage}\label{sec:leakage}

Here we explore how the locally varying Overhauser fields from Eq.~(\ref{eq:HF_coupled}) cause leakage outside the qubit Hilbert space with different settings of external magnetic fields and exchange couplings. This leakage originates from the quasistatic nuclear spin dynamics (relative to electron spin one), and averaging such quasistatic dynamics leads to coherence decays on the inhomogeneous dephasing time scale $T_2^*$.\cite{Merkulov_PRB02} We are thus motivated to discuss treatments to remove the $T^{*}_{2}$ decays in the current section.

\subsubsection{Noninteracting spins with zero applied magnetic field: $\EZ=J=0$}\label{sec:J0EZ0} \label{sec:noninteracting}
When both the exchange coupling $J$ and the external magnetic field $E_Z$ vanish, all eight electron spin eigenstates of the triple QD are degenerate.  The electron spin dynamics is then dominated by the random Overhauser fields described in Eq.~(\ref{eq:HF_coupled}).  The logical qubit states, $\ket{D^{\prime}_{1/2}}$ and $\ket{D_{1/2}}$, are hf-coupled to each other and to all the other states except $\ket{Q_{-3/2}}$, and are not stationary.
If the system is initialized into a superposition of $\ket{D^{\prime}_{1/2}}$ and $\ket{D_{1/2}}$, it would evolve into the full 8-dimensional basis over the time scale of inhomogeneous broadening dephasing $T_2^*$. 
In principle, spin echo methods could be used to undo this leakage, due to the quasistatic nature of the Overhauser fields. However, since the instantaneous Overhauser fields change in time, the quantization axes for the electron spins are unknown, making it impossible to perform simple spin echoes that rotate the system back to its initial state.

\subsubsection{Noninteracting spins in a finite magnetic field: $J=0$ and $\EZ\neq 0$}\label{sec:J0}
When an external field is applied, but $J=0$, the energy spectrum splits into the $S^z$ manifold structure shown in Fig.~\ref{fig:3spinSpectrumCoupled}(b). The logical qubit belongs to the $S^z$=$1/2$ manifold, which includes the states $\ket{D^{\prime}_{1/2}}$, $\ket{D_{1/2}}$, and $\ket{Q_{1/2}}$. After subtracting $\frac{1}{2}(E_Z + B_{010}^z)\mathbb{1}$, the lowest-order Hamiltonian in this 3-dimensional subspace is given by
\begin{equation}
\hat{H}^{S^{z}=1/2}_{\text{hf}} \!=\! \left(\!
\begin{array}{ccc}
	\frac{1}{3} B_{1\bar{2}1}^z \!&\! -\frac{1}{2\sqrt{3}} B_{10\bar{1}}^z
			\!&\! -\frac{1}{3\sqrt{2}} B_{1\bar{2}1}^z \\
	-\frac{1}{2\sqrt{3}} B_{10\bar{1}}^z \!&\! 0
			\!&\! -\frac{1}{\sqrt{6}} B_{10\bar{1}}^z \\
	-\frac{1}{3\sqrt{2}} B_{1\bar{2}1}^z \!&\! -\frac{1}{\sqrt{6}} B_{10\bar{1}}^z
			\!&\! \frac{1}{6} {B}_{1\bar{2}1}^z
\end{array}\!
\right)\label{eq:H025}.
\end{equation}
Clearly, the three-level dynamics is determined by the two Overhauser field gradients, $B_{1\bar{2}1}^z$ and $B_{10\bar{1}}^z$.  The subspace is fully coupled, so an initial superposition between $\ket{D^{\prime}_{1/2}}$ and $\ket{D_{1/2}}$ would eventually leak into state $\ket{Q_{1/2}}$.  Moreover, assuming the Overhauser fields are randomized, one can easily show that $\langle B_{1\bar{2}1}^z B_{10\bar{1}}^z \rangle = 0$; i.e., the two gradient terms are uncorrelated.  Consequently, the leakage rates for $\ket{D^{\prime}_{1/2}}$ and $\ket{D_{1/2}}$ are uncorrelated, leading to quick state mixing that seems difficult to reverse in this three-level subspace, since the electron spin quantization axes are determined by the instantaneous but random nuclear fields.

When $J = 0$, the three spins are not coupled.  In the absence of hf interaction and in the presence of an external magnetic field, the three-spin eigenstates split into four Zeeman manifolds, with the $S^z = 1/2$ manifold containing states $\ket{\uparrow \uparrow \downarrow}$, $\ket{\uparrow \downarrow \uparrow}$, and $\ket{\downarrow \uparrow \uparrow}$.  The hf interaction takes a particularly simple form in this product state basis:
\begin{equation}
\hat{H}^{\rm product}_{\text{hf}} \!=\! \frac{1}{2}\left(\!
\begin{array}{ccc}
	B_{11\bar{1}}^z & 0 & 0 \\
	0 & B_{1\bar{1}1}^z & 0 \\
	0 & 0 & {B}_{\bar{1}11}^z
\end{array}\!
\right)\label{eq:product}.
\end{equation}
Since these product states are not directly coupled by the hf interaction at the lowest order, we can express the time evolution of $\ket{D^{\prime}_{1/2}}$, $\ket{D_{1/2}}$, and $\ket{Q_{1/2}}$ by
\begin{eqnarray}
\!\!\!\!\ket{D^{\prime}_{1/2}} &\!\!=\!\!& \frac{1}{\sqrt{3}}
	\left(|\uparrow \uparrow \downarrow\rangle - 2e^{-iB_{0\bar{1}1}^z t}|\uparrow \downarrow \uparrow \rangle + e^{-iB_{\bar{1}01}^z t} |\downarrow \uparrow \uparrow\rangle\right), \quad\quad\\
\!\!\!\!\ket{D_{1/2}} &\!\!=\!\!& \frac{1}{\sqrt{2}}
	\left(-|\uparrow \uparrow \downarrow\rangle + e^{-iB_{\bar{1}01}^z t}|\downarrow \uparrow \uparrow \rangle \right),\\
\!\!\!\!\ket{Q_{1/2}} &\!\!=\!\!& \frac{1}{\sqrt{3}} \left(|\uparrow \uparrow \downarrow\rangle + e^{-iB_{0\bar{1}1}^z t}|\uparrow \downarrow \uparrow \rangle + e^{-iB_{\bar{1}01}^z t}|\downarrow \uparrow \uparrow\rangle \right) .
\end{eqnarray}
Thus $\ket{D^{\prime}_{1/2}}$ and $\ket{D_{1/2}}$ would quickly acquire additional phases in the superpositions because of the nuclear fields. Without spin echo, the dephasing times for $\ket{D^{\prime}_{1/2}}$ and $\ket{D_{1/2}}$ would be determined by the nuclear field gradients, in the order of $T_2^*$, where $T_2^*$ is the single-spin dephasing time.  On the positive side, the quantization axis for each individual electron is well defined with the finite external field and the vanishing coupling between electron spins, and spin echoes can be performed on each of the spins.  To correct the effects of the random Overhauser fields on the three-spin product states, we need to perform simultaneous spin echoes on each of the individual electron spins.  Such simultaneous echoes will not only recover individual product states, but also allow superposition states between $|e\rangle$ and $|g\rangle$ to survive beyond $T_2^*$.

\subsubsection{Exchange-coupled spin chain in zero magnetic field: $\EZ=0$ and $J\neq0$}\label{sec:EZ0}
Figure~\ref{fig:3spinSpectrumCoupled}(a) shows the relevant energy manifolds when $E_Z\ll J$.
The quadruplet states $Q$ are split off from the logical qubit states, and they do not have a strong effect on the qubit evolution.  The logical qubit states $\ket{D^{\prime}_{1/2}}$ and $\ket{D_{1/2}}$ are not isolated, however; they each have a Zeeman companion state.  Therefore, the logical qubit is part of two distinct manifolds: $D^{\prime}$ and $D$.  The intra-manifold degeneracy is broken by the small random nuclear fields.  The hf interaction also provides a coupling between the manifolds. The coupling is small however, when $B^z_d\ll J$. In this limit, we can focus on the intra-manifold dynamics.  The lowest-order Hamiltonians in the respective manifolds are given by
\begin{eqnarray}
\hat{H}^{D^{\prime}} & = &  \left(
	\begin{array}{cc}
		-J + \frac{1}{6} B_{2\bar{1}2}^z & -\frac{1}{6} B_{2\bar{1}2}^-  \\
		-\frac{1}{6} B_{2\bar{1}2}^+ & -J - \frac{1}{6} B_{2\bar{1}2}^z
	\end{array}\right)  \;,\label{eq:H01}\\
\hat{H}^{D} & = &  \left(
	\begin{array}{cc}
		\frac{1}{2} B_{010}^z & -\frac{1}{2} B_{010}^-  \\
			-\frac{1}{2} B_{010}^+ & -\frac{1}{2} B_{010}^z
	\end{array}\right)\;.
\end{eqnarray}
Thus the random nuclear polarization in the QDs drives random rotations within each of the manifolds, and leads to leakages from both the $\ket{D^{\prime}_{1/2}}$ and $\ket{D_{1/2}}$ states.

As an example, we calculate the inhomogeneous broadening due to the random $z$-axis Overhauser fields $B^{z}_{d}$ in the $D^{\prime}$ manifold, neglecting the effects of a transverse Overhauser field. As analyzed in Ref.~\onlinecite{Merkulov_PRB02}, such random Overhauser fields are quasistatic over the electron spin precession period. We assume that $B^{z}_{d}$ follows a Gaussian distribution $P(B^{z}_{d})= \frac{1}{\sqrt{2\pi} \sigma_{z,d}}e^{-(B^{z}_{d}-\bar{B}^{z}_{d})^{2}/2\sigma^{2}_{z,d}}$, with a mean value $\bar{B}^{z}_{d}$ and a variance $\sigma^{2}_{z,d}$.
By averaging over the phase difference $e^{-i\frac{1}{3}B^{z}_{2\bar{1}2}t}$ between the two $D^{\prime}$ states, we obtain the inhomogeneous broadening decay,
\begin{eqnarray}
	Q_{D^{\prime}}(t) &\equiv& \langle e^{-i\frac{1}{3}B^{z}_{2\bar{1}2}t}\rangle_{\text{en}}
		=\prod^{3}_{d=1}\int dB^{z}_{d}\, P(B^{z}_{d}) e^{-iq_{d}B^{z}_{d}t} \nonumber\\
		&=& e^{-\frac{1}{18}(4\sigma^{2}_{z,1}+\sigma^{2}_{z,2}+4\sigma^{2}_{z,3})t^{2}}e^{-i\frac{1}{3}\bar{B}^{z}_{2\bar{1}2}t}  \,, \label{eq:Q01}
\end{eqnarray}
where $\langle\cdots\rangle_{\text{en}}$ denotes an ensemble average over the bath, and $q_{d}$ takes the values $2/3,-1/3,2/3$ for $d=1,2,3$, respectively.

\begin{figure}[t]
	\includegraphics[width=0.8\linewidth]{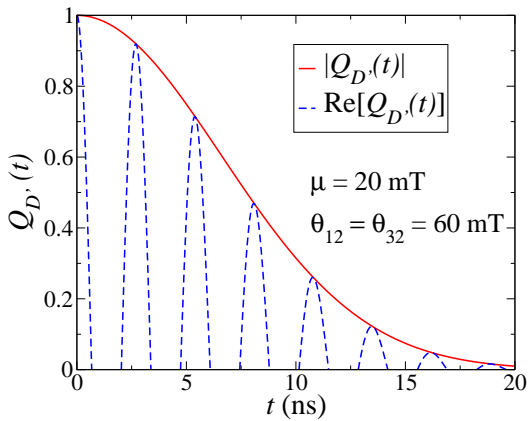}
\caption{Possible qubit leakage dynamics $Q_{D^{\prime}}(t)$ in GaAs (red solid line).
Both curves assume the same field gradients.
(See the main text for the definitions of $\mu$, $\theta_{12}$ and $\theta_{32}$.)
The blue dashed curve corresponds to the case of a mean gradient $\bar{B}^{z}_{2\bar{1}2}\neq 0$,\cite{Foletti_NP09} while the red curve corresponds to $\bar{B}^{z}_{2\bar{1}2}= 0$.}\label{fig:leakage}
\end{figure}

In Eq.~(\ref{eq:Q01}), we see that the characteristic decay time scale, $T^{*}_{D^{\prime}}=3\sqrt{2}/(4\sigma^{2}_{z,1}+\sigma^{2}_{z,2}+4\sigma^{2}_{z,3})^{1/2}$, is comparable to the time scale for inhomogeneous broadening in a single spin, $T^{*}_{2}$.  For example, if the dots are identical, so that $\sigma_{z,d}=\sigma_{z,s}$ ($d=1,2,3$), then we obtain $T^{*}_{D^{\prime}}=\sqrt{2}/\sigma_{z,s}$.  (Here, $\sigma_{z,s}$ is the standard deviation of the longitudinal Overhauser field for a single QD.)
When $\bar{B}^{z}_{2\bar{1}2}\neq 0$, an extra phase $e^{-i\frac{1}{3}\bar{B}^{z}_{2\bar{1}2}t}$ appears in $Q_{D^{\prime}}(t)$.
In this case, the oscillations represent the coherent leakage dynamics, while the decay envelope describes the inhomogeneous broadening.
It may be easier to create a controlled field gradient between the dots (e.g., by dynamic nuclear polarization\cite{Foletti_NP09}), rather than a constant polarization in a single dot.  We may therefore express $\frac{1}{3}\bar{B}^{z}_{2\bar{1}2}$ explicitly in terms of field gradients: $\frac{1}{3}\bar{B}^{z}_{2\bar{1}2}=\bar{\mu}+\frac{1}{3}\bar{\theta}_{12}+\frac{1}{3}\bar{\theta}_{32}$.
Here, $\bar{\mu}\!=\!(\bar{B}^{z}_{1}\!+\!\bar{B}^{z}_{2}\!+\!\bar{B}^{z}_{3})/3$ is the average longitudinal Overhauser field, while $\bar{\theta}_{12}=\bar{B}^{z}_{1}\!-\!\bar{B}^{z}_{2}$ and $\bar{\theta}_{32}=\bar{B}^{z}_{3}\!-\!\bar{B}^{z}_{2}$ correspond to the Overhauser field gradients across QDs $1$ and $2$, and QDs $3$ and $2$, respectively.
In Fig.~\ref{fig:leakage}, we plot $Q_{D^{\prime}}(t)$ as a function of time for a triple QD in GaAs with random nuclear fields, both with and without a mean gradient $\bar{B}^{z}_{2\bar{1}2}$.

If the nuclear spins are not specially prepared (via dynamic nuclear polarization, for example), the random Overhauser fields yield electron spin quantization axes that are time-dependent
for the three electrons.  Similar to the case of $J=0$ and $E_Z = 0$ in Sec.~\ref{sec:noninteracting}, dephasing due to random Overhauser fields cannot then be corrected by spin echoes.  If a finite $\bar{B}^{z}_{2\bar{1}2}$
is present and known (e.g., by dynamic nuclear polarization), so that the quantization axis for $\ket{D^{\prime}_{1/2}}$ and $\ket{D^{\prime}_{-1/2}}$ is clearly defined, then a spin echo experiment may be possible.  However, performing a spin echo in this manifold requires a pulsed field gradient in the form of $B^{\pm}_{2\bar{1}2}$, which is technically challenging to generate.  Moreover, the required pulses would defeat the main purpose of three-spin encoding:  to enable all-electrical control without pulsed magnetic fields.
We therefore conclude that the absence of a uniform magnetic field makes the three-spin logical qubit vulnerable to inhomogeneous broadening.

\subsubsection{Preferred operating regime for reducing leakage}\label{sec:operating_regime}

In the three scenarios we have studied so far,  random nuclear spin polarizations cause leakage to the outside of the logical qubit space.  When nuclear configurations are quasistatic, it is often (though not always) possible to employ spin-echo or more sophisticated dynamical decoupling techniques to reverse the leakage effects. However, these manipulations inevitably require controlled flips of the physical spins (as opposed to logical qubits).  As discussed in the previous section, these operations with fast magnetic control are difficult to perform for our encoding scheme, and should be avoided if at all possible.

To fight leakage while maintaining fully electrical control, a simple method would be to require $E_Z,J \gg \sigma_{z,d}$ (hf fluctuations) at all times.
In this case, leakage predominantly involves just a single eigenstate ($\ket{Q_{1/2}}$), and even this leakage is suppressed for large $J$.

For the original exchange-only qubit proposed in Ref.~\onlinecite{DiVincenzo_Nature00}, the exchange interaction is turned off while the qubits are idling.  Our previous discussion indicates that the logical qubit would be vulnerable to local magnetic noise from hf interactions during such idle periods.  It would therefore be challenging to realize the proposed qubits in materials with finite hyperfine interaction, such as GaAs, and to a lesser degree, natural Si.  A possible solution could be to recalibrate the pulse sequences, with a constant, nonzero $J$ between neighboring dots.   A recent experimental implementation of the resonant-exchange-qubit employs such constant couplings in a triple QD.\cite{Medford_PRL13}  In this arrangement, the condition $J,E_Z \gg \sigma_{z,d}$ is satisfied and leakage is suppressed.  In the remainder of this paper, we focus on this parameter regime.

\subsection{Effective Hamiltonian for pure dephasing}\label{sec:Heff}
We now consider the regime $E_Z, J\gg \sigma_{z,d}$, where the dominant leakage state ($\ket{Q_{1/2}}$) is separated from the logical qubit states $\ket{g} =\ket{D^{\prime}_{1/2}}$ and $\ket{e} = \ket{D_{1/2}}$ by the large, controllable exchange splitting $J/2$, and the remaining three-spin eigenstates are split off by an even larger Zeeman energy. In this regime, hf interactions cannot cause direct transitions between any of the three-spin eigenstates.  Their leading-order effect is a second-order modification of the energies of the states, thus causing dephasing. Relaxation via spin-orbit interactions is also very weak because it requires two spin flips. The dominant decoherence mechanism arising from hf interactions is therefore pure dephasing within the logical qubit subspace.

To study this behavior, we first construct an effective Hamiltonian in the two-dimensional ($\ket{D^{\prime}_{1/2}}$-$\ket{D_{1/2}}$) subspace. We perform a Schrieffer-Wolff transformation to decouple $\ket{D^{\prime}_{1/2}}$ and $\ket{D_{1/2}}$ from the remaining Hilbert space,\cite{Winkler} taking hf interactions as the perturbation.  We then obtain
\begin{eqnarray}
	\hat{H}_{\text{eff}} &\!\!=&\!\! \left( \begin{array}{cc}
		\!-J+\hat{H}_{A} + \hat{V} + \hat{H}_{\text{N}} & 0 \!\\\! 0 & -\hat{H}_{A} - \hat{V} + \hat{H}_{\text{N}} \!\end{array} \right). \quad\label{eq:Heff}
\end{eqnarray}
Here, we have only retained leading-order terms that dominate the dephasing physics under various conditions, as explained below. Furthermore, we impose the condition $\EZ \gg J$. As we will show later, this condition yields a simple mathematical form in the dephasing terms, and limits the number of dephasing channels.

In the $\EZ\gg J$ limit, the $\hat{H}_{A}$ and $\hat{V}$ terms in Eq.~(\ref{eq:Heff}) can be written as hf couplings, summed over the individual QDs:
\begin{equation}
	\hat{H}_{A} = \sum^{3}_{d=1} \nu_{d} \hat{H}_{A,d} \;, \;
		\hat{V} = \sum^{3}_{d=1}\nu_{d} \hat{V}_{d}\;. \label{eq:HA_V}
\end{equation}
Here, $\hat{H}_{A,d}=\sum_{i\in d} \frac{A_{i}}{2}I^{z}_{i}$ is first order in the longitudinal hf fields, while $\hat{V}_{d}=\sum_{i,j \in d} \frac{A_{i}A_{j}}{4\EZ} I^{-}_{i} I^{+}_{j}$ is second order in the transverse hf fields.
$\hat{V}_d$ represents the leading-order coupling between different Zeeman manifolds, as indicated by the Zeeman energy denominator.
Note that analogous terms arise in the leading-order hf couplings for single-spin qubits.\cite{Cywinski_PRB09}
In Eq.~(\ref{eq:HA_V}), the $\nu_{d}$ coefficients describe the spin distribution in the different QDs, as given in Table~\ref{table:coupled_spin_states}.  For logical qubit states, they take the values $\nu_{d}=1/3,-2/3,1/3$, for $d=1,2,3$, respectively.

In the effective Hamiltonian $\hat{H}_{\rm eff}$, we have omitted off-diagonal terms of order $\langle \hat{H}_{A} \rangle$.  When $J\gg \sigma_{z,d}$, these terms perturb the energy at order $\langle \hat{H}_{A} \rangle^2/J \ll \langle \hat{H}_{A} \rangle$.  They also induce relaxation processes at the same order.  The off-diagonal terms are therefore much weaker than the $\hat{H}_A$ terms along the diagonal.  We note that the same considerations apply to effective Hamiltonians involving the main leakage state ($\ket{Q_{1/2}}$).

The nuclear spin flip-flop term $\hat{V}$ in $\hat{H}_{\rm eff}$ contains exclusively intradot nuclear spin flipflops as indicated in Eq.~(\ref{eq:HA_V}), because our focus here is on the noninteracting limit $\EZ\gg J$.
The electron spins are thus dephased by the locally varying nuclear fields in each dot. The current situation in the noninteracting limit is analogous to $S$-$T_{0}$ qubits in the far-detuned regime.\cite{Petta_Science05}  In the latter case, the two electrons are spatially separated, and dephasing is caused by the locally varying nuclear fields. Nuclear spin flip-flops involving \emph{two} QDs arise at first order in $J/E_Z$.  This higher order effect is relatively weak to those included in Eq.~(\ref{eq:Heff}), and we do not consider it here.  However, we provide some details of the calculation in the Appendix.

The effective Hamiltonian $\hat{H}_{\rm eff}$ of Eq.~(\ref{eq:Heff}) has the same mathematical structure as the dephasing Hamiltonian for a single electron spin.\cite{Cywinski_PRB09}   The only significant difference is that the Zeeman splitting $E_Z$ for a single spin is now replaced by the exchange splitting $J$ in $\hat{H}_{\text{eff}}$. This replacement makes the exchange-only logical qubit directly sensitive to charge noise, since the exchange splitting $J$ originates from electrical interactions.  We will explore the charge decoherence channel elsewhere; here we focus on hyperfine interactions.

The similarities between $\hat{H}_{\rm eff}$ for the logical qubit from Eq.~(\ref{eq:Heff}) and the single-spin dephasing Hamiltonian in Ref.~\onlinecite{Cywinski_PRB09} allows us to glean useful insights into logical qubit decoherence.  It is instructive to first recall the basic physics of single-spin dephasing due to hyperfine interactions.  At typical experimental temperatures ($\sim$~100~mK) and fields ($<$10~T), the nuclear spins are in the effective high-temperature limit,\cite{Hanson_RMP07} and their orientations are random.  Additionally, since the nuclear magneton is three orders of magnitude smaller than the Bohr magneton, the nuclear dynamics is effectively quasistatic compared to the electron spins.\cite{Merkulov_PRB02, Taylor_PRB07}  The leading effect of the nuclear spins on the electron dynamics is therefore inhomogeneous broadening, due to the slowly varying Overhauser fields.  In GaAs this $T_2^*$ time scale is on the order of 10~ns.\cite{Merkulov_PRB02, Petta_Science05}  If the nuclear spin distribution can be narrowed, so the longitudinal Overhauser field is constant in time and the lowest-order inhomogeneous broadening is suppressed, the transverse Overhauser field can still lead to dephasing of the electron spin at the next order.  In GaAs, this time scale is on the order of 1~$\mu$s.\cite{Cywinski_PRB09}  Both of these dephasing effects can be corrected by spin echo techniques.  At the next-leading-order, dephasing is governed by interference between the different nuclear spin species.  In GaAs, this $T_2$ time scale is on the order of tens of $\mu$s.\cite{Bluhm_NP11}

By analogy, for three-spin logical qubits we expect that inhomogeneous broadening due to longitudinal Overhauser fields should be the main nuclear-induced dephasing mechanism.  If the longitudinal Overhauser field can be made constant, the logical qubit will still experience dephasing from transverse Overhauser fields in the form of a narrowed-state free induction decay. Similar to single spins, this inhomogeneous broadening can be removed, to leading order, by applying a Hahn spin echo, or a more complicated pulse.  The most important distinction with single-spin qubits is that pulses in three-spin qubits are all-electrical.  For example, $x$ rotations can be implemented via an ac modulation of $\Delta =J_{23} - J_{12}$ in Eq.~(\ref{eq:He}), at the resonant frequency of the logical qubit.\cite{Medford_PRL13}

\subsection{Pure dephasing dynamics of the logical qubit}\label{sec:dephasing}

Our focus now is on nuclear-spin-induced dephasing during free evolution of the logical qubit, or the free induction decay (FID).  This decoherence, in analogy to the single-spin-qubit case,\cite{Cywinski_PRB09} is described by
\begin{equation}
	W^{\text{FID}}(t) = \Tr_{I} \left(\hat{\rho}_{I} e^{it\hat{H}_{22}} e^{-it\hat{H}_{00}} \right) \label{eq:FID}\;.
\end{equation}
where $\hat{\rho}_{I}$ is the nuclear density operator, and $\hat{H}_{00}$ ($\hat{H}_{22}$) represents the effective nuclear spin Hamiltonian when the electron spins are in $\ket{D^{\prime}_{1/2}}$ ($\ket{D_{1/2}}$).

The FID dynamics can be probed by Ramsey-fringe type experiments, in which two $\pi/2$ pulses with a time interval $t$ in between are applied on the qubit initially prepared in an eigenstate.  For example, if the qubit is initially in the $|0\rangle$ state.  A first $\pi/2$ pulse about the $x$ axis brings the qubit to a superposition state $\frac{1}{\sqrt{2}}(\ket{D^{\prime}_{1/2}}-i\ket{D_{1/2}})$.  After the qubit freely evolves for a time period $t$, a second $\pi/2$ pulse is applied.  Assuming the pulses are instantaneous, the projected probability of the ground state after the second pulse is given by $P_{g}(t)=\frac{1}{2}+\frac{1}{2}\text{Re}[W^{\text{FID}}(t)]$.  The evolution of $W^{\text{FID}}(t)$ can thus be tracked.

Below we evaluate the qubit dephasing by calculating $W^{\text{FID}}(t)$ based on Hamiltonian~(\ref{eq:Heff}).  Specifically, we first determine the inhomogeneous dephasing time scale $T^{*}_{2}$ from the quasistatic longitudinal Overhauser field $\hat{H}_A$, then calculate the narrowed-state FID time scale $T^{\text{FID}}$ due to the transverse Overhauser field (contained in $\hat{V}$) by the approach described in Ref.~\onlinecite{Cywinski_PRB09}.  Since the inhomogeneous broadening induced dephasing can be removed by a Hahn echo (HE) or other pulse sequences, we further calculate the dephasing dynamics with the application of a HE and CPMG pulse sequence.

\subsubsection{Inhomogeneous broadening}

We first investigate the inhomogeneous broadening of the logical qubit due to the longitudinal Overhauser field $\hat{H}_A$.  Qualitatively, fluctuations in $\hat{H}_A$ cause dephasing between the two qubit states.  Since $\hat{H}_A$ is quasistatic, $2\hat{H}_A$ can be written as a classical Overhauser field $\sum_d \nu_d B_{d}^z$.   Under typical experimental conditions the nuclear spins are at the high-temperature limit, their orientations completely random.  As such, the net Overhauser field on each dot is in a Gaussian distribution around 0, and this distribution leads to an inhomogeneous broadening in the qubit energy splitting through $\hat{H}_A$.  Following an analysis similar to Eq.~(\ref{eq:Q01}), we find that $\hat{H}_A$ contributes the following dephasing factor to  $W^{\text{FID}}(t)$:
\begin{eqnarray}
	W^{*}(t) & \equiv & \Tr_{I} \left(\hat{\rho}_{I} e^{-2it\hat{H}_A} \right) = \prod^{3}_{d=1} \int dB^{z}_{d} P(B^{z}_{d}) \ e^{-i\nu_{d}B^{z}_{d} t}\nonumber\\
		&=& \exp{\left(-\frac{1}{2}\sum^{3}_{d=1}\nu^{2}_{d}\sigma^{2}_{z,d}t^{2}\right)} \;. \label{eq:W*}	
\end{eqnarray}
The resulting inhomogeneous broadening dephasing time is
\begin{equation}
	T^{*}_{2}=\frac{3\sqrt{2}}{\sqrt{\sigma^{2}_{z,1}+4\sigma^{2}_{z,2}+\sigma^{2}_{z,3}}} \sim \frac{\sqrt{3}}{\sigma_{z,s}}\;,
\end{equation}	
where $\sigma_{z,s}$ is the standard deviation of the longitudinal Overhauser field in a single-spin QD.

Clearly, the inhomogeneous broadening dephasing time $T^{*}_{2}$ here is roughly the same as that for a single spin qubit in a single QD.  The slight difference comes from the altered distribution of the single-spin-density.
	
\subsubsection{The Narrowed-State Free Induction Decay}

In this section we consider qubit dephasing by a ``properly narrowed'' nuclear bath, which is determined by $\hat{V}$.  Such a nuclear bath has a narrowed superposition of $\hat{H}_A$ eigenstates, so that the corresponding standard deviations $\sigma_{z,d}$ for $\hat{H}_A$ in the triple dot are much smaller than their respective thermal state values.
Consequently, the inhomogeneous broadening associated with $\hat{H}_A$ is strongly suppressed. At the limit of complete suppression, $\hat{H}_{A}$ is a constant, and does not cause any decoherence to the logical qubit, so that the lowest order qubit decoherence is caused by $\hat{V}$.
Experimentally, various techniques have been developed to achieve nuclear state narrowing and reduce the width of $\sigma_{z,d}$ or the standard deviation of an Overhauser field gradient across a double QD ($\propto\sigma_{z,d}$).\cite{Greilich_Science06, Greilich_Science07, Reilly_Science08, Bluhm_PRL10}  In the rest of the paper, we will refer to this properly narrowed FID as ``nFID''.

The nuclear spin flipflops in $\hat{V}$ are mediated by the electron spins in the logical qubit.  The resulting random dynamics in the nuclear spin reservoir in turn leads to phase fluctuations and thus pure dephasing in the logical qubit.  The nFID from this flip-flop channel has been calculated for a single-spin QD in Ref.~\onlinecite{Cywinski_PRB09} using a ring diagram theory, in which the authors perturbatively expand $\hat{V}_{d}|_{\nu_{d}=1}$ and resum the linked-cluster terms from the expansion. The resulting nFID for single-spin qubits is a nonexponential decay, with a phase shift.

Here we adapt the ring diagram technique to calculate logical qubit dephasing (specifically nFID) due to $\hat{V}$.  Since $\hat{V}$ contains three commuting flip-flop channels from the three dots, the nFID for the qubit can be factored into a product of contributions $W^{\text{nFID}}_{d}(t)$ from each dot,
\begin{eqnarray}
	\!W^{\text{nFID}}(t) &\!\!=\!\!& \prod^{3}_{d=1} W^{\text{nFID}}_{d}(t) \quad\label{eq:FID} \\
		&\!\!=\!\!& \prod^{3}_{d=1} \Tr_{I}\left\{\hat{\rho}^{I}\bar{\mathcal{T}}
			\left[e^{i\nu_{d}\int^{t}_{0}\mathcal{V}_{d}(\tau)d\tau}\right]\!
			\mathcal{T}\left[e^{-i\nu_{d}\int^{t}_{0}\mathcal{V}_{d}(\tau)d\tau}\right]\right\}\nonumber,
\end{eqnarray}
where $\mathcal{T}$ ($\bar{\mathcal{T}}$) is the time (anti-)ordering operator, and $\mathcal{V}_{d}(\tau)\!=\!\nu_{d}\sum_{k,l\in d}\frac{A_{k}A_{l}}{4\EZ}I^{-}_{k}I^{+}_{l}e^{-i(\omega_{k[\alpha]}-\omega_{l[\beta]})t}$ represents $\hat{V}_{d}$ in the interaction picture with respect to $\hat{H}_{\text{N}}$.  For the nuclear spin splitting we do not include the Knight shift from the hf interaction.  Such an approximation only causes a quantitative error for times $t> N/\MA$ (e.g., in GaAs, with $\MA\approx 130$ $\mu$eV and the number of unit cells per dot $N=10^{6}$, $N/\MA \sim 30$ $\mu$s), making it valid for all the calculations presented in this paper.  We also note that the nFID of a single spin in a single dot is given by $W^{\text{s,nFID}}(t) = \left. W^{\text{nFID}}_{d}(t) \right|_{\nu_{d}=1}$.

Now that the nFID of the logical qubit is expressed in terms of single-spin nFIDs in single dots (with the interaction strengths modified by factor $\nu_d$), we can use the existing result of a single dot in Ref.~\onlinecite{Cywinski_PRB09} and obtain
\begin{equation}
	W^{\text{nFID}}_{d}(t) = \prod_{m} \frac{e^{-i\arctan{\lambda^{d}_{m}(t)}}}{\sqrt{1+\lambda^{d}_{m}(t)^{2}}}\label{eq:W_d}\;,
\end{equation}
with $\lambda^{d}_{m}(t)$ being the eigenvalues of the coarse-grained matrix for dot $d$
\begin{equation}
T^{d}_{\alpha\beta}(t)=\nu_{d}\sqrt{a_{\alpha}a_{\beta}}\sqrt{n_{\alpha}n_{\beta}} \frac{\MA_{\alpha}\MA_{\beta}}{N\EZ} e^{-i\frac{\omega_{\alpha\beta}t}{2}}\frac{\sin{(\frac{\omega_{\alpha\beta}t}{2})}}{\omega_{\alpha\beta}} \,.
	\label{eq:Tkl}
\end{equation}
Here $a_{\alpha}\equiv\langle I^{-}_{\alpha}I^{+}_{\alpha}\rangle$ at zero nuclear polarization and $n_{\alpha}$ denotes the nuclear concentration of the species $\alpha$. $\MA_{\alpha}$ and $\omega_{\alpha\beta}\!=\!\omega_{\alpha}\!-\!\omega_{\beta}$ are the hf interaction strength of the nuclear species $\alpha$, and  the Zeeman energy difference between nuclear species $\alpha$ and $\beta$, respectively.  Here we have assumed three roughly identical dots, such that the number of nuclear spins in each dot is about the same, i.e., $N_d = N$.  This matrix $T^{d}_{\alpha\beta}(t)$ has the dimension of $N_{\alpha}$, the number of nuclear species, and its diagonal (off-diagonal) elements are responsible for the contribution from the flip-flop process of homonuclear (heteronuclear) spin pairs.

\begin{figure}[t]
	\includegraphics[width=0.9\linewidth]{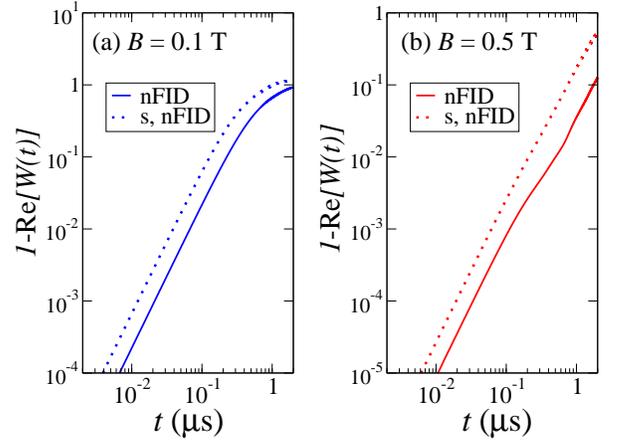}
\caption{The coherence loss $\{1-\text{Re}[W^{\text{nFID}}(t)]\}$ (solid lines) and $\{1-\text{Re}[W^{\text{s,nFID}}(t)]\}$ (dotted lines) as a function of time, in log-log scale, based on Eqs.~(\ref{eq:FID}) and (\ref{eq:W_d}) for GaAs at (a) $B=0.1$ T and (b) $B=0.5$ T.}\label{fig:loglog}
\end{figure}

\begin{figure}[t]
	\includegraphics[width=0.9\linewidth]{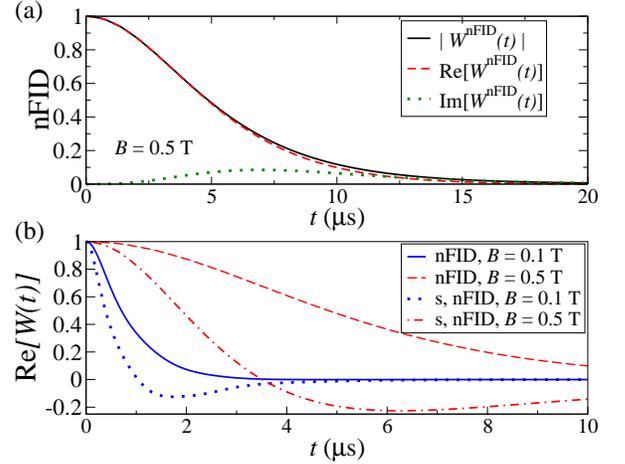}
\caption{The nFID coherence function as a function of time, based on Eqs.~(\ref{eq:FID}) and (\ref{eq:W_d}) for GaAs. (a) $|W^{\text{nFID}}(t)|$ (black solid line), $\text{Re}[W^{\text{nFID}}(t)]$ (red dashed line) and $\text{Im}[W^{\text{nFID}}(t)]$ (green dotted line) at $B=0.5$ T with $N=10^{6}$. (b) The comparison of $\text{Re}[W(t)]$ between the encoding and single-spin qubit made at $B=0.1$ T (blue solid line for the nFID and blue dotted line for the single-spin nFID) and at $B=0.5$ T (red dashed line for the nFID and red dash-dotted line for the single-spin nFID), with $N=10^{6}$.}\label{fig:fid_gaas}
\end{figure}

Figures~\ref{fig:loglog} and ~\ref{fig:fid_gaas} present numerical results obtained from Eqs.~(\ref{eq:FID}) and (\ref{eq:W_d}) on the time-dependence of the qubit coherence function. In these simulations, we set the nuclear concentration $n_{\alpha}=0.6$, $0.4$, and $1.0$ for $^{69}$Ga, $^{71}$Ga and $^{75}$As, respectively. Figure~\ref{fig:loglog} focuses on the short-time behavior by plotting the results on a log-log scale, while Fig.~\ref{fig:fid_gaas} gives a better representation of the longer-time behavior of the coherence function.

To better understand the qualitative behaviors of the various curves presented in Figs.~\ref{fig:loglog} and ~\ref{fig:fid_gaas}, we discuss two regimes of time in more detail: (1) The short-time limit, when $t\!\ll\!\omega^{-1}_{\alpha\beta}$. At this limit, nuclear spins can be considered having identical precession frequencies, and the expression~(\ref{eq:Tkl}) becomes a scalar: $T^{d}_{\alpha\alpha}(t)=\nu_{d}\eta t$, with $\eta\equiv n_{\alpha}a_{\alpha}\MA^{2}_{\alpha}/(2N\EZ)$. (2) A Longer-time limit, when $t\!\gg\!\omega^{-1}_{\alpha\beta}$ (but still short enough so that Knight-shift effects are not prominent).  At such a time scale the nuclear spins tend to flip-flop with only those of the same species, so that $T^{d}_{\alpha\beta}(t)$ is diagonal.  For GaAs, with three nuclear species, the time scale $\omega^{-1}_{\alpha\beta}$ is roughly 3 $\mu$s at $B = 0.1$ T.

At the short-time limit $t\ll\omega^{-1}_{\alpha\beta}$ (or in a dot with a single nuclear species), the single-dot contribution to nFID, $W^{\text{nFID}}_{d}(t)$, takes the form\cite{Cywinski_PRB09}
\begin{equation}
	W^{\text{nFID}}_{d} \left(t\ll \omega^{-1}_{\alpha\beta} \right) \approx \frac{e^{-i\arctan{ (\nu_{d} \eta t)}}}{\sqrt{1+ (\nu_{d} \eta t)^{2}}}\label{eq:W_dcl}\,,
\end{equation}
which is characterized by the time scale $(\nu_{d}\eta)^{-1}$ that accounts for the strength of the spin flip-flop channel $\nu_{d}\hat{V}_{d}$. This expression~(\ref{eq:W_dcl}) can also be obtained semiclassically, by considering $\hat{V}_{d}$ as the mean field of a large number of randomized nuclear spins, and calculating the coherence factor $e^{-2i\nu_{d}V_{d}t}$.\cite{Neder_PRB11}

The short-time nFID for the logical qubit now takes the form
\begin{eqnarray}
	\!\!\!\!\!&&W^{\text{nFID}} \left(t\ll \omega^{-1}_{\alpha\beta} \right) = e^{-i\sum^{3}_{d=1}\arctan{ (\nu_{d} \eta t)}}\label{eq:W_expand}\\
	\!\!\!\!\!&&\quad\nonumber \times \left[1+\eta^{2}t^{2} \sum^{3}_{d=1} \nu^{2}_{d}
	+ \eta^{4}t^{4} \sum_{d\neq d'} \nu^{2}_{d}\nu^{2}_{d'}
	+ \eta^{6}t^{6}\prod^{3}_{d=1} \nu^{2}_{d} \right]^{-1/2}.
\end{eqnarray}
The denominator of $W^{\text{nFID}}(t\!\ll\!\omega^{-1}_{\alpha\beta})$ contains both quadratic and higher-order terms in $t$ because it is a product of three $W^{\text{nFID}}_{d}(t)$. This is different from the single-spin case, which is given by\cite{Cywinski_PRB09, Neder_PRB11}
\begin{equation}
	W^{\text{s,nFID}} \left(t\ll \omega^{-1}_{\alpha\beta}\right) =
		\frac{e^{-i\arctan{ (\eta t)}}}{\sqrt{1+(\eta t)^{2}}}\label{eq:W_single}.
\end{equation}
When $\eta t\gtrapprox 1$ (or $\hat{V}$ has its full effect), Eq.~(\ref{eq:W_expand}) indicates that the three-spin nFID should be faster than the single-spin nFID because of the higher-order terms in time.

The product form of the three-spin FID also leads to a reduced phase shift in $W^{\text{nFID}}(t)$ as compared with the single-spin nFID. Equation~(\ref{eq:W_expand}) has a phase factor $e^{-i\sum^{3}_{d=1}\arctan{ (\nu_{d} \eta t)}}$. When $\hat{V}$ is just starting to dephase the qubit i.e., $\eta t\!\ll\!1$, the total phase shift can be approximated by $-i\eta t \sum^{3}_{d=1}\nu_{d}\!=\!0$.  As $t$ increases, the total phase shift in Eq.~(\ref{eq:W_expand}) may still be small, because the transverse Overhauser field $\nu_{2}\hat{V}_{2}$ in dot 2 is always out of phase with those from dots 1 and 3, leading to a partial phase cancellation.  On the other hand, in the single-spin case, the phase shift $-i\eta t$ is finite and grows with time linearly.  If a physically observable quantity is proportional to Re$[W^{\rm nFID}(t)]$, significant differences in its dynamics could arise from this difference in phase shift between the logical qubit and a single-spin qubit.  Nevertheless, it is important to note here that the phase shift is not a loss of coherence, but a modification to the qubit energy splitting from the nuclear spin flip-flop processes; what destroys the qubit coherence is the nonexponential decay in Eq.~(\ref{eq:W_expand}).

Figure~\ref{fig:loglog} focuses on the initial behavior of the nFID by plotting the coherence loss $\{1-\text{Re}[W^{\text{nFID}(t)}]\}$ and $\{1-\text{Re}[W^{\text{s,nFID}(t)}]\}$ on a log-log scale, based on Eqs.~(\ref{eq:FID}) and (\ref{eq:W_d}). The simulations are made for GaAs at $B\!=\!0.1$ T and $0.5$ T. The data of $B\!=\!0.1$ T in Fig.~\ref{fig:loglog}(a) are within the $t\ll\omega^{-1}_{\alpha\beta}$ regime, while in Fig.~\ref{fig:loglog}(b), the larger magnetic field reduces the $\omega^{-1}_{\alpha\beta}$ ($\sim 0.6$ $\mu$s), so that the $t \ll \omega^{-1}_{\alpha\beta}$ covers only the initial segment of the curves. The slopes of the lines in Fig.~\ref{fig:loglog} at small $t$ are approximately $2$, indicating that the terms $\propto t^{2}$ dominate at such short times.

To characterize the nFID described in Eq.~(\ref{eq:W_expand}), we define a characteristic time $T^{\text{nFID}}$ when $|W(T^{\text{nFID}})|=e^{-1}$ is satisfied. Given the values of $\nu_{d}$, we find numerically
\begin{equation}
	T^{\text{nFID}} \approx \frac{2.2}{ \eta} \approx 0.88\, T^{\text{s,nFID}}  \label{eq:T2_short}\;,
\end{equation}
where $T^{\text{s,nFID}}$ is the characteristic time of $W^{\text{s,nFID}}(t\ll \omega^{-1}_{\alpha\beta})$ for a single spin defined in an analogous manner.  The factor of 0.88 comes from the higher-than-quadratic terms contained in Eq.~(\ref{eq:W_expand}), which shows that the effect of these terms is quite mild. Using Eq.~(\ref{eq:T2_short}), we estimate that for a GaAs triple dot with $N=10^{6}$, $T^{\text{nFID}}\sim 0.9$ $\mu$s for $B=0.1$ T, and $\sim 5$ $\mu$s for $B=0.5$ T.

When $t\gg\omega^{-1}_{\alpha\beta}$ (but still $t\!<\!N/\MA$), the expression~(\ref{eq:Tkl}) takes on a diagonal form: $T^{d}_{\alpha\beta}(t)\!\approx\!\delta_{\alpha \alpha^{\prime}} T^{d}_{\alpha \alpha^{\prime}}(t)$, and the $d$ dot nFID function can be expressed as a product of contributions from each of the nuclear species: $W^{\text{nFID}}_{d}(t)\!\approx\!\prod^{N_{\alpha}}_{\alpha}W^{\text{nFID}}_{d,\alpha}(t)$.  The $\alpha$-species contribution $W^{\text{nFID}}_{d,\alpha}(t)$ takes the form
\begin{equation}
	W^{\text{nFID}}_{d,\alpha}(t \gg \omega^{-1}_{\alpha\beta}) = \frac{e^{-i\arctan{ (\nu_{d}\eta_{\alpha} t)}}}{\sqrt{1+(\nu_{d}\eta_{\alpha} t)^{2}}},\label{eq:W_dalpha}
\end{equation}
with $\eta_{\alpha}\equiv n_{\alpha}\eta$. The approximated nFID for the logical qubit is then given by
\begin{equation}
	W^{\text{nFID}}(t \gg \omega^{-1}_{\alpha\beta}) = \prod^{3}_{d=1}\prod^{N_{\alpha}}_{\alpha}W^{\text{nFID}}_{d,\alpha}(t \gg \omega^{-1}_{\alpha\beta})\;. \label{eq:FID_long}
\end{equation}

We emphasize again that we focus on the $t<N/\MA$ regime (e.g., $t < 30$ $\mu$s for GaAs with $N=10^{6}$), and the $t\gg \omega^{-1}_{\alpha\beta}$ case considered here is not a true long-time limit. For the nFID at times $t>N/\MA$, the Knight field effect should be included, and one should also consider other dephasing mechanisms such as nuclear dipole-dipole interaction.

In Fig.~\ref{fig:fid_gaas}, we present the results for the nFID in a GaAs triple dot at two different magnetic fields, based on Eqs.~(\ref{eq:FID}) and (\ref{eq:W_d}). Here our focus is the nFID behavior after the $t\ll\omega^{-1}_{\alpha\beta}$ period. The $W^{\text{nFID}}(t)$ function at $B\!=\!0.5$ T is plotted in Fig.~\ref{fig:fid_gaas}(a). From the $|W^{\text{nFID}}(t)|$ curve, we find $T^{\text{nFID}}\!\approx\!6.5$ $\mu$s different from the estimate by Eq.~(\ref{eq:T2_short}), because the $\omega^{-1}_{\alpha\beta}\ll 1$ approximation no longer applies to the case of $B=0.5$ T. [In the case of $B=0.1$ T, Eq.~(\ref{eq:T2_short}) yields an estimate of $T^{\text{nFID}}$ with the discrepancy $\sim10\%$.] In Fig.~\ref{fig:fid_gaas}(a), the small and slowly varying imaginary part $\text{Im}[W^{\text{nFID}}(t)]$ is the consequence of the reduced phase shift for a three-spin logical qubit.
In Fig.~\ref{fig:fid_gaas}(b), we compare the three-spin nFID with the single-spin nFID (in terms of $\text{Re}[W(t)]$, which is measurable) at $B=0.1$ T and $B=0.5$ T.  The dramatic differences between the single-spin curves and the logical qubit curves are mostly due to the finite phase factors for the single spin.

In summary, we find that the narrowed-state free induction decay due to hyperfine interaction for the logical qubit encoded in a triple dot is qualitatively similar to that for a single-spin qubit in a single quantum dot, except a near cancellation of the phase shift due to hyperfine interaction.  The total coherence function can be factored into a product of contributions from the three dots at the $J \ll E_Z$ limit, making the physical picture of three-spin decoherence a simple analogy to the single-spin case.

\subsubsection{Spin echo and CMPG pulse sequences}

It is well known that dynamic pulse sequences such as the Hahn and CPMG pulse sequences can completely remove the inhomogeneous broadening effect from the quasistatic longitudinal Overhauser field\cite{Slichter} represented by $\hat{H}_{A}$. In this section, we calculate the dynamics of a Hahn echo (HE) and a two-pulse CPMG echo for the logical qubit.

Generally, we find the echo decays calculated below have the the same oscillatory behavior as what has been found for a single-spin qubit,\cite{Cywinski_PRL09, Cywinski_PRB09,Bluhm_NP11} because of the similarities in the effective Hamiltonians and the nFID of the logical qubit and a single spin qubit. This oscillatory behavior originates from the different Larmor frequencies of the nuclear spin species in the host material, which lead to beatings in the transverse Overhauser field in $\hat{V}$. For a QD formed by a single species, this oscillation is absent, so that one has to consider other dephasing processes that may cause spin echo decay.  Strictly speaking, this beating is not exactly a true decay of a spin echo.  The echo signal can recover almost fully as the different beatings synchronize again after certain periods of time.  However, when additional interactions are taken into consideration, whether nuclear dipolar coupling or higher-order nuclear spin flipflops due to hyperfine interaction,\cite{Cywinski_PRB09} the re-synchronization will not be complete.  With the high-fidelity requirement imposed on qubits, we can thus focus on the initial loss of coherence.

To produce an HE, a bit-flip pulse is applied after a qubit is put in a superposed state and evolves freely for a period of $t/2$.  The qubit coherence at time $t$ then takes the form
\begin{eqnarray}
	W^{\text{HE}}(t) &=& \Tr_{I} \left( \hat{\rho}_{I} e^{i\hat{H}_{22}t/2 }e^{i\hat{H}_{00}t/2 }
		e^{-i\hat{H}_{22}t/2}e^{-i\hat{H}_{00}t/2} \right) \nonumber\\
		&=& \Tr_{I}\left\{ \hat{\rho}_{I} \mathcal{T}_{C}\left[e^{-i\int_{C}f^{\text{HE}}(\tau_{c})\mathcal{V}_{c}(\tau_{c})d\tau_{c}} \right] \right\}\,,\label{eq:HE}
\end{eqnarray}
where $\mathcal{V}_{c}(\tau_{c})$ denotes the time-contour form of $\hat{V}$ in the interaction picture,  operated with the time-contour ordering operator $\mathcal{T}_{C}$ and the filter function $f^{\text{HE}}(t)$ for HE defined in Ref.~\onlinecite{Cywinski_PRB09}.

\begin{figure}[t]
	\includegraphics[width=0.85\linewidth]{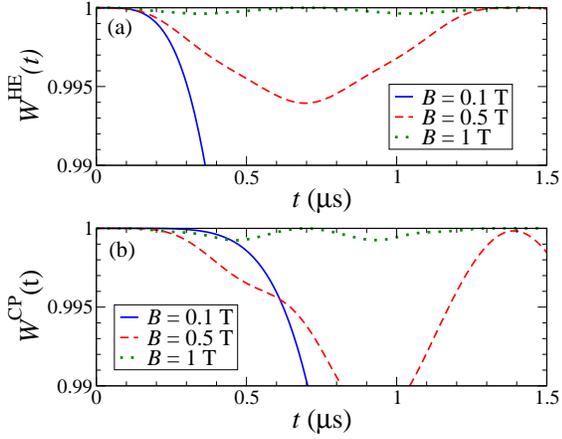}
	\caption{The coherence functions with the application of (a) Hahn echo and (b) 2-pulse CPMG sequence, for GaAs with $N=10^{6}$ at $B=0.1$ T (blue solid line), $B=0.5$ T (red dashed line) and $B=1$ T (green dotted line). In both panels, $N=10^{6}$.}
\label{fig:se_cp2_gaas}
\end{figure}
\begin{figure}[t]
	\includegraphics[width=0.85\linewidth]{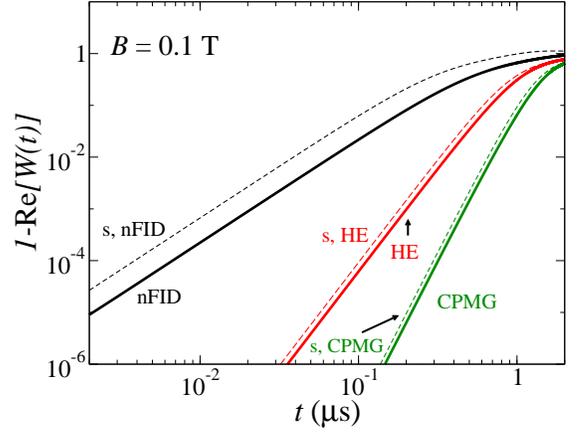}
	\caption{The coherence loss $\{1-\text{Re}[W(t)]\}$ for a GaAs triple dot at $B=0.1$ T in the case of nFID (black solid line, with the smallest slope) and with the application of a Hahn echo pulse (red solid line) and 2-pulse CPMG sequence (green solid line, with the largest slope). The dashed lines correspond to the relevant coherence loss for a single-spin qubit. We set $N=10^{6}$ for all data.}
\label{fig:dd_loglog}
\end{figure}

Similarly to the calculation for nFID, the HE decay for the logical qubit is given by $W^{\text{HE}}(t)=\prod^{3}_{d=1}W^{\text{HE}}_{d}(t)$, with $W^{\text{HE}}_{d}(t)$ representing the HE decay in QD $d$. Again, the HE decay in the single-spin case is given by $W^{\text{s,HE}}(t)=W^{\text{HE}}_{d}(t)|_{\nu_{d}=1}$.
We consider the lowest-order contribution in time for the HE decay in Eq.~(\ref{eq:HE}) (since short-time dynamics is more relevant), and obtain\cite{Cywinski_PRB09}
\begin{equation}
	W^{\text{HE}}_{d}(t)=\frac{1}{ 1+\frac{1}{2}|\nu_{d}|^{2} R^{\text{HE}}(t)}\;,\label{eq:W^HE}
\end{equation}
where $|\nu^{2}_{d}|R^{\text{HE}}(t)$ corresponds to the lowest-order term from the perturbative expansion with\cite{Hung_PRB13}
\begin{equation}
	R^{\text{HE}}(t)\!=\! 4\sum_{\alpha\neq\beta} a_{\alpha}a_{\beta}n_{\alpha}n_{\beta} \left(\frac{\MA_{\alpha}\MA_{\beta}}{N\EZ}\right)^{2} \frac{\sin^{4}{(\frac{\omega_{\alpha\beta}t}{4})}}{\omega^{2}_{\alpha\beta}}. \label{eq:R^HE}
\end{equation}
The term $|\nu_{d}|^{2}$ in Eq.~(\ref{eq:W^HE}) modifies $W^{\text{HE}}_{d}(t)$ quantitatively.  However, the HE decay mainly depends on the behavior of $R^{\text{HE}}(t)$.

To better understand the behavior of $R^{\text{HE}}(t)$, we rewrite Eq.~(\ref{eq:R^HE}) as $R^{\text{HE}}(t)\!=\!\sum\!R^{\text{HE}}_{\alpha\beta}[1- \cos{ (\frac{\omega_{\alpha \beta}t}{2})}]^{2}\!>\!0$, with the magnitude $R^{\text{HE}}_{\alpha\beta} \propto (E_{Z} \omega_{\alpha \beta})^{-2} \propto B^{-4}$.  For a triple dot (with $\MA_{\alpha}$, $n_{\alpha}$ and $N$ fixed), the HE decay $W^{\text{HE}}(t)$ in a low magnetic field tends to have a slower oscillating frequency ($\omega_{\alpha\beta}\propto B$) and a larger amplitude ($R^{\text{HE}}_{\alpha\beta}\propto B^{-4}$).
These two trends are clearly illustrated in Fig. \ref{fig:se_cp2_gaas}(a), where we present the simulations of $W^{\text{HE}}(t)$ for GaAs at $B=0.1$ T, $B=0.5$ T and $B=1$ T. The $W^{\text{HE}}(t)$ function at $B=0.1$ T shows a larger initial decay than those at the other magnetic fields, and it recovers at a longer time.

We also calculate the coherence decay when a 2-pulse CPMG sequence is applied by the same approach.\cite{Cywinski_PRB09} This pulse sequence contains two $\pi$ pulses at $t/4$ and $3t/4$, respectively, and the echo appears at time $t$. The resulting coherence decay $W^{\text{CP}(t)}$ takes the form
\begin{eqnarray}
	W^{\text{CP}}(t) &=& \Tr_{I}\left\{ \hat{\rho}_{I} \mathcal{T}_{C}\left[e^{-i\int_{C}f^{\text{CP}}(\tau_{c})\mathcal{V}_{c}(\tau_{c})d\tau_{c}} \right] \right\}\\
	&=& \prod^{3}_{d=1} \frac{1}{1 + \frac{1}{2} |\nu_{d}|^{2} R^{\text{CP}}(t)}\,.
\end{eqnarray}
where $f^{\text{CP}}(t)$ is the filter function for 2-pulse CPMG given in Ref.~\onlinecite{Cywinski_PRB09}, and
the $R^{\text{CP}}(t)$ function is given by
\begin{equation}
 	\!R^{\text{CP}}(t)\!=\! 16\sum_{\alpha\neq\beta}\! a_{\alpha}a_{\beta}n_{\alpha}n_{\beta}\!
		\left(\! \frac{\MA_{\alpha} \MA_{\beta}}{N\EZ} \! \right)^{2} \! \frac{\sin^{2}{(\frac{\omega_{\alpha\beta}t}{4})} \sin^{4}{( \frac{\omega_{\alpha\beta}t}{8})}}{\omega^{2}_{\alpha\beta}}. \label{eq:R^CP}
\end{equation}
The $R^{\text{CP}}(t)$ function, compared with $R^{\text{HE}}(t)$, contains more oscillatory terms, and brings a richer behavior to $W^{CP}(t)$.
For the terms contributing to $R^{\text{CP}}(t)$ and $R^{\text{HE}}(t)$ with the same $\omega_{\alpha\beta}$, we find the ratio $R^{\text{HE}}_{\alpha \beta}(t) / R^{\text{CP}}_{\alpha \beta}(t)\!=\!\cot^{2}{\left(\frac{\omega_{\alpha\beta}t}{8}\right)}$.  Thus, initially $R^{\text{HE}}(t)\!>\!R^{\text{CP}}(t)$, and the 2-pulse CPMG sequence has a better performance than the HE sequence, until $\cot^{2}{(\frac{\omega_{\alpha\beta}}{8}T_{R})}\!=\!1$.  In Fig.~\ref{fig:se_cp2_gaas}(b), we show our calculated $W^{\text{CP}}(t)$ for GaAs at $B=0.1$ T, $B=0.5$ T and $B=1$ T.  Clearly, the initial decay is slower in panel (b) than in panel (a) for each of the given fields, especially at lower fields.

To compare the performance of HE and CPMG, we plot in Fig.~\ref{fig:dd_loglog} the initial loss of coherence $\{1-\text{Re}[W(t)]\}$ for GaAs in the case of FID, and with the application of a HE and a 2-pulse CPMG sequence. As expected, the application of the CPMG pulse sequence yields a better performance in maintaining coherence than HE, which in turn improves over the narrowed-state FID.  We note that the curves for FID, HE, and CPMG all have different slopes, representing an initial $t^2$, $t^4$, and $t^6$ dependence, respectively.  These time dependencies are consistent with previous studies on dynamical decoupling for single-electron spin qubits,\cite{Cywinski_PRB09} and demonstrate again the progressively improving coherence from narrowed state FID to CPMG echo.  In addition, the difference between the two nFID curves seems to be larger than those for HE and CPMG.  This is due to the finite phase shift for the single-spin nFID as compared to the nearly vanishing phase shift for the logical qubit that we have discussed in the nFID section.  For HE and CPMG the coherence functions are real and do not have the extra phase shift.

In short, for the logical qubit, the Hahn echo and CPMG echo dynamics due to hyperfine interaction with nuclear spins are very similar to the single-spin cases, not only qualitatively but also quantitatively.

\section{Three-Spin Decoherence with Nonuniform Coupling}\label{sec:nonuniform}

So far we have focused on decoherence properties of a triple dot with uniform exchange coupling: $J_{12} = J_{23}$.  However, one of the most important advantages to the three-spin encoding is that it can be completely controlled electrically via the two exchange couplings.  It is thus inevitable that the triple dot would spend time in regimes where $J_{12} \neq J_{23}$.  In this section we explore the coherence properties of a nonuniform chain.

\subsection{Effective Hamiltonian}
As we have discussed in Sec.~\ref{sec:3spins}, the effect of the nonuniformity in exchange coupling leads to coupling between $\ket{D^{\prime}_{1/2}}$ and $\ket{D_{1/2}}$ (and $\ket{D^{\prime}_{-1/2}}$ and $\ket{D_{-1/2}}$), and energy shifts in most of the uniform chain states, but most importantly, no coupling between the $\ket{D^{\prime}_{1/2}}-\ket{D_{1/2}}$ manifold with any other three-spin states.  As such we can still use $\ket{D^{\prime}_{1/2}}$ and $\ket{D_{1/2}}$ or their superpositions as states of a logical qubit.  In a finite magnetic field, the Hamiltonian projected on the $\ket{D^{\prime}_{1/2}}-\ket{D_{1/2}}$ basis now takes the form
\begin{equation}
	\!\!\!\!\hat{H}_{\Delta}\!=\!\left(\!\!\begin{array}{cc}
			-J -\frac{\Delta}{2} & \frac{\sqrt{3}\Delta}{4} \\
			\frac{\sqrt{3}\Delta}{4}  & 0 \end{array} \!\!\right) +
		\left(\!\!\begin{array}{cc}
			\hat{H}_{A}\!+\!\hat{V}\!+\!\hat{H}_{N}  & \frac{-B^{z}_{10\bar{1}}}{2\sqrt{3}}\!+\!\hat{V}^{ge}\\
			\frac{-B^{z}_{10\bar{1}}}{2\sqrt{3}}\!+\!\hat{V}^{eg}  & -\hat{H}_{A}\!-\!\hat{V}\!+\!\hat{H}_{N}\end{array} \!\!\right)	
		\label{eq:H^0-2_NU}
\end{equation}
Here we have included the off-diagonal terms that were omitted in Hamiltonian~(\ref{eq:Heff}), and $\hat{V}^{ge}$ in the $\EZ \gg J$ limit is given by $\hat{V}^{ge} = \hat{V}^{eg} = \frac{-1}{\sqrt{3}} (\hat{V}_{1}-\hat{V}_{3})$.
Both $\hat{V}$ and $\hat{V}^{ge}$ contain only intradot nuclear spin flip-flops because we consider the $\EZ \gg J$ regime. \emph{Interdot} nuclear spin flipflops between nearest-neighbor QDs would appear if we kept terms $\propto J/E_Z$, and all interdot flipflops become possible if we include terms linear in $\Delta/E_Z$.  In other words, in the cases when the system does not have any symmetry and the Zeeman splitting from the applied magnetic field is not much larger than the exchange couplings, all possible nuclear spin flip-flops can be mediated by the electrons, and they can contribute to the three-electron-spin dephasing.

Hamiltonian (\ref{eq:H^0-2_NU}) for the logical qubit in a nonuniform triple dot is formally similar to the effective Hamiltonian for a singlet-triplet qubit in the presence of a magnetic field gradient.\cite{Hung_PRB13}  The nonuniformity in the exchange coupling, $\Delta$, plays the same role as the magnetic field gradient in the case of a double dot.  As we have done in Ref.~\onlinecite{Hung_PRB13}, here we first diagonalize the electronic part in Hamiltonian (\ref{eq:He_NU}), and obtain the eigenstates $\ket{g^{\prime}}=\cos{\theta} \ket{g} + \sin{\theta} \ket{e}$ and $\ket{e^{\prime}}= -\sin{\theta} \ket{g} + \cos{\theta} \ket{e}$, where the rotation angle $\theta$ is given by $\tan{2\theta}\equiv -\frac{\sqrt{3}\Delta}{2}(J+\frac{\Delta}{2})^{-1}$.  We thus obtain the unitary transformation that rotates the hf terms of Hamiltonian (\ref{eq:H^0-2_NU}) to the new eigenbasis and makes it a pure dephasing Hamiltonian.

The total effective Hamiltonian diagonal in the $\ket{g^{\prime}}-\ket{e^{\prime}}$ basis (effects of the leftover off-diagonal terms are negligible) is
\begin{eqnarray}
	\hat{H}_{\text{eff},\Delta} = \left(\begin{array}{cc}
		E^{\prime}_{g} + \hat{H}^{\prime}_{A} + \hat{V}^{\prime} & 0 \\
		0 & E^{\prime}_{e} - \hat{H}^{\prime}_{A} - \hat{V}^{\prime} \end{array}\right)\;,\label{eq:Heff_NU}
\end{eqnarray}
where $E^{\prime}_{g,e}= \frac{-1}{2}(J+\frac{\Delta}{2})(1\pm \sec{2\theta})$, $\hat{H}^{\prime}_{A} = \sum^{3}_{d=1} \nu^{\prime}_{d}\hat{H}_{A,d}$, and $\hat{V}^{\prime} = \sum^{3}_{d=1} \nu^{\prime}_{d}\hat{V}_{d}$.  The values of $\nu^{\prime}_{d}$, which are determined by the electron spin density in each of the dots, are given by
\begin{eqnarray}
	\nu^{\prime}_{1} &=& \frac{\cos{2\theta}-\sqrt{3}\sin{2\theta}}{3} = \frac{2}{3} \cos \left(2\theta + \frac{\pi}{3} \right)\;, \nonumber \\
    \nu^{\prime}_{2} & = & -\frac{2}{3} \cos{2\theta} \;,\nonumber \\
	\nu^{\prime}_{3} &=& \frac{\cos{2\theta}+\sqrt{3}\sin{2\theta}}{3} = \frac{2}{3} \cos \left(2\theta - \frac{\pi}{3} \right)\;.\label{eq:nu_prime}
\end{eqnarray}
When $\theta \rightarrow 0$, $\nu^\prime_d$ goes back to $\nu_d$.  Equation (\ref{eq:nu_prime}) shows that the nonuniformity $\Delta=J_{23}-J_{12}$ introduces a quantitative change to the logical qubit: the electron spin density has been redistributed over the triple dot based on the rotation angle $\theta$. When $\Delta$ vanishes, $\theta=0$, and $\hat{H}_{\text{eff},\Delta}$ recovers the form of $\hat{H}_{\text{eff}}$ in Sec.~\ref{sec:decoherence}. We also note that $\nu^{\prime}_{1} \neq \nu^{\prime}_{3}$ when $\Delta\neq 0$, since the $S_{13}$ symmetry is broken.  While one can choose a $\Delta$ such that $\nu^{\prime}_{1}$ (or $\nu^{\prime}_{3}$) vanishes, such a $\Delta$ makes the Overhauser field in the other QDs stronger simultaneously.

\subsection{Coherence decay with nonuniform coupling}

Structurally $\hat{H}_{\text{eff},\Delta}$ is identical to $\hat{H}_{\text{eff}}$; thus we expect the qubit dephasing dynamics to be quite similar between uniform and nonuniform triple dots.  Physically this is entirely reasonable, since for the logical qubit states in a nonuniform triple dot, the total spin remains $S=1/2$, and spin polarization remains $S_z = 1/2$.  The only change is that the already-finite spin density is redistributed among the three dots.

We first consider the inhomogeneous broadening effects of $\hat{H}^\prime_A$.  Following a similar analysis to that in Eq.~(\ref{eq:W*}), we obtain dephasing time $T^*_{2,\Delta}$ due to inhomogeneous broadening from $\hat{H}^\prime_A$ as
\begin{equation}
	T^{*}_{2,\Delta} = \sqrt{\frac{2}{\sum^{3}_{d=1}(\nu^{\prime}_{d})^{2}\sigma^{2}_{z,d}}}\sim \frac{1}{\sigma_{z,s}}\;.
\end{equation}
Indeed, since $\sum^{3}_{d=1}(\nu^{\prime}_{d})^{2} = \sum^{3}_{d=1}(\nu_{d})^{2} = 2/3$, we find $T^{*}_{2,\Delta}$ similar to that in a uniformly coupled triple dot, which falls on the order of single-spin $T^{*}_{2}$. The generally small modifications are consistent with the qualitative discussion presented in the previous subsection.

As for the nFID due to $\hat{V}^{\prime}$, we can directly use the expression $W^{\text{nFID}}_{d}(t)$ in Eq.~(\ref{eq:W_d}), replacing $\nu_{d}$ by $\nu^{\prime}_{d}$, to calculate $W^{\text{nFID}}(t)=\prod^{3}_{d=1}W^{\text{nFID}}_{d}(t)$.
\begin{figure}[t]
	\includegraphics[width=0.85\linewidth]{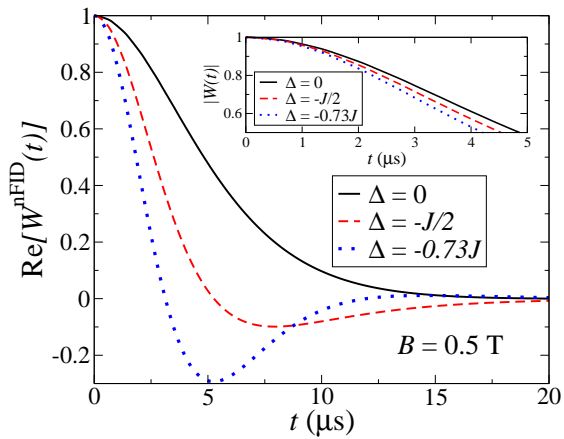}
	\caption{$\text{Re}[W^{\text{nFID}}(t)]$ with $\Delta=0$ (black solid line), $\Delta= -J/2$ (red dashed line), and $\Delta= -0.73 J$ (blue dotted line) at $B=0.5$ T for GaAs with $N=10^{6}$. These $\Delta$ values correspond to the cases of $J_{23}=J_{12}$, $J_{23}=J_{12}/2$, and $J_{23}=0.27J_{12}$. The inset shows $|W^{\text{nFID}}(t)|$ at short times for these three $\Delta$ values, which are very similar.}
\label{fig:W_NU}
\end{figure}
At times $t \ll \omega^{-1}_{\alpha\beta}$, the nFID takes the form of Eq.~(\ref{eq:W_expand}) with $\nu_{d}$ replaced by $\nu^{\prime}_{d}$. Since $\sum_{d}\nu'^{2}_{d}=\sum_{d}\nu^{2}_{d}=2/3$, the initial nFID here, if the phase term contained in Eq.~(\ref{eq:W_expand}) were absent, would be almost identical to the $\Delta=0$ case.

In Fig.~\ref{fig:W_NU}, we plot $\text{Re}[W^{\text{nFID}}(t)]$ at three different values of $\Delta$ for GaAs at $B=0.5$ T.  The $\Delta=0$ line in Fig.~\ref{fig:W_NU} is provided as a benchmark, while the results with $\Delta=-J/2$ and $\Delta=(1-\sqrt{3})J \approx -0.73 J$ correspond to the situation where one of the Overhauser fields is vanishing ($\nu^{\prime}_{1}=0$) and the case of $\tan{2\theta}=1$. As expected, the logical qubits in nonuniform triple dots dephase in times that are quite similar to a qubit in a uniform triple dot.

\section{Conclusions}\label{sec:conclusion}

In this paper we study hyperfine induced decoherence of three-electron-spin states in a semiconductor triple quantum dot, specifically the states involved in the three-spin encoding in the ($S=1/2,S^{z}=1/2$) subspace.  We first delineate the full three-spin Hilbert space (without any orbital excitations) and obtain the electronic eigenstates.  We then write the hyperfine interaction Hamiltonian in this eight-state basis, which allows us to explore how different three-spin states are coupled and modified by the hyperfine interaction with the nuclear spins.

To establish the viability of the three-spin logical qubit from the perspective of decoherence, we first focus on its leakage, and identify $E_{Z}, J \gg \sigma_{z,d}$ as the regime where information leakage via the hyperfine interaction can be effectively suppressed. In such regime, we construct the effective pure dephasing Hamiltonian in the $E_{Z} \gg J$ limit.  When qubit leakage is suppressed, the main decoherence mechanisms for the logical qubit are charge noise and hyperfine interaction with nuclear spins, and we focus on the latter in this work.

We calculate logical qubit dephasing due to the random longitudinal Overhauser field, which causes inhomogeneous broadening and free induction decay.
We also examine pure dephasing due to the random transverse Overhauser field, or nuclear spin flip-flops, which lead to the narrowed-state free induction decay.  Lastly, we calculate decay of the Hahn echo and the 2-pulse CPMG echo.  For these calculations we either take the semiclassical approach (to deal with the quasistatic longitudinal Overhauser field) or use the ring diagram theory, with parameters for typical GaAs quantum dots.  We find that all the relevant decoherence times are of the same order as those for a single-spin qubit.  The modifications compared to the single-spin case are due to the electron spin density spreading over the three dots, which also leads to faster decay at longer times.

When the logical qubit is operated in the regime where $J_{12}\neq J_{23}$, the $\Delta=J_{23}-J_{12}$ makes the qubit Hilbert space rotated, but no leakage from the qubit space.  The qubit rotation causes a redistribution of the spin density in the triple dot.  However, since the total spin density remains the same, the nonuniformity in the exchange coupling does not qualitatively change the dephasing dynamics as compared to the case of a uniform triple dot.

In short, our calculations show that a three-spin encoded logical qubit using the $(S=1/2, S^z = 1/2)$ states has very similar decoherence properties to those of a single-spin qubit when finite exchange coupling is constantly employed to suppress qubit leakage. However, considering that constantly-on exchange coupling is required for this logical qubit to operate properly, its decoherence properties due to charge noise \cite{Coish_PRB05, Hu_PRL06, Culcer_APL13} and electron-phonon interaction \cite{Hu_PRB11, Gamble_PRB12} should also be clarified.

The authors acknowledge financial support by US ARO (W911NF0910393) and NSF PIF (PHY-1104672 and PHY-1104660). This work was supported by the United States Department of Defense. The US government requires publication of the following disclaimer: the views and conclusions contained in this document are those of the authors and should not be interpreted as representing the official policies, either expressly or implied, of the US Government.

\appendix
\section{THE LOWEST-ORDER CORRECTION IN $J/\EZ$ TO THE EFFECTIVE HAMILTONIAN}\label{app:J/E_Z}

As we have discussed in Sec.~\ref{sec:Heff}, the effective Hamiltonian (\ref{eq:Heff}) contains only intradot nuclear spin flip-flops because we focus on the $J \ll E_Z$ regime.  When $J$ is larger, so that terms linear in $J/\EZ$ should be kept in the spin Hamiltonian, interdot dephasing channels will appear in the expressions of $\hat{V}$, $\hat{V}^{ge}$, and $\hat{V}^{eg}$.  Here we clarify their form and magnitude.

We define $\delta\!\equiv\!J/\EZ$, and write $\hat{V}\!=\!\hat{V}(0)\!+\!\hat{V}(\delta)$, where $\hat{V}(0)$ is independent of $\delta$ and given in Eq.~(\ref{eq:HA_V}). The correction $\hat{V}(\delta)$ includes an intradot flip-flop process in dot $2$ and an interdot flip-flop channel for a pair of nuclear spins from adjacent dots denoted by $\hat{V}_{d_{1},d_{2}}$:
\begin{equation}
	\hat{V}(\delta) = \frac{\delta}{6}\left( 5\hat{V}_{2}- 2 \hat{V}_{d_{1},d_{2}} \right),\label{eq:V_r}
\end{equation}
where
\begin{equation}
	\hat{V}_{d_{1},d_{2}}= \sum_{i \in d_{1}, j \in d_{2}}\frac{A_{i} A_{j}}{4\EZ} I^{-}_{i}I^{+}_{j} \,.
\end{equation}

In the case of a nonuniform triple dot (when $\Delta \neq 0$), we write $\hat{V}^{ge}\!=\!\hat{V}^{ge}(0)+ \hat{V}^{ge}(\delta)$ and $\hat{V}^{eg}\!=\!\hat{V}^{ge}(0)+\hat{V}^{eg}(\delta)$, where $\hat{V}^{ge}(0)\!=\!\frac{-1}{\sqrt{3}}(\hat{V}_{1}-\hat{V}_{3})$, and $\hat{V}^{ge}(\delta)$ and $\hat{V}^{eg}(\delta)$ take the form
\begin{eqnarray}
	\hat{V}^{ge}(\delta) &\!\!=\!\!& \frac{\delta}{\sqrt{3}}\left( \hat{V}_{1}-\hat{V}_{3}
		- \frac{\hat{V}_{2,1} - \hat{V}_{2,3} + 3 \hat{V}_{1,2} - 3\hat{V}_{3,2}}{2}\right),\qquad\label{eq:V02_r}\\
	\hat{V}^{eg}(\delta) &\!\!=\!\!&  \frac{\delta}{\sqrt{3}}\left(\hat{V}_{1}-\hat{V}_{3} - \frac{\hat{V}_{1,2} - \hat{V}_{3,2} + 3 \hat{V}_{2,1} - 3 \hat{V}_{2,3}}{2}\right).\qquad\label{eq:V20_r}	
\end{eqnarray}
Note that $\hat{V}_{d_{1},d_{2}}$ and $\hat{V}_{d_{2},d_{1}}$ are distinct flip-flop processes.

\bibliography{ref_qd}
\end{document}